\documentstyle{article}
\begin{document}
\begin{titlepage}
YCTP-P10-92

\smallskip
UTTG-07-92

\bigskip

\today

\bigskip
\centerline{\bf{$S$ AND $T$ MATRICES FOR THE SUPER $U(1,1)$ WZW MODEL.}}
\centerline{\bf{APPLICATION TO SURGERY AND 3-MANIFOLDS INVARIANTS}}
\centerline{\bf{BASED ON THE ALEXANDER CONWAY POLYNOMIAL}}

\bigskip

\centerline{L.Rozansky\footnote{Research supported in part by the Robert
A.Welch Foundation and NSF Grant PHY9009850}}
\centerline{Physics Department}
\centerline{Texas University at Austin}
\centerline{Austin TX 78712 USA}

\bigskip

\centerline{H.Saleur\footnote{On leave from SPHT Cen Saclay 91191 Gif Sur
Yvette Cedex France. Work supported by DOE Contract number DE-AC02-76ERO3075
 and by a Packard Fellowship
for Science and Engineering}}
\centerline{Physics Department}
\centerline{Yale University}
\centerline{New Haven CT06511 USA}

\bigskip

{\bf Abstract}: We  carry on (in a self
contained fashion) the study of
 the Alexander Conway
invariant from the quantum field theory point of view started in \cite{RS91}.
We investigate for that purpose various aspects of WZW models on supergroups.

 We first discuss in details
$S$ and $T$ matrices for the $U(1,1)$ super WZW model and obtain, for the
 level $k$
an integer, new finite dimensional representations of the modular group. These
 have
the remarkable property that some of the $S$ matrix elements are infinite (we
show how to properly handle such divergences). Moreover, typical and  atypical
representations as well as
indecomposable blocks are mixed: truncation to maximally atypical
 representations,
as advocated  in some recent papers, is not consistent.

 Using our approach,
multivariable Alexander invariants for links in $S^{3}$
can now be fully computed by surgery. Examples of torus and
 cable knots are
discussed. Consistency with classical results provides independent checks of
the solution of the $U(1,1)$ WZW model.

The main topological application of this work is the computation of Alexander
invariants for 3-manifolds and more generally for links in 3-manifolds.
 Invariants of 3-manifolds themshelves
seem to depend trivially on the level $k$,
but still contain interesting topological information. For Seifert manifolds
for instance, they essentially coincide with the order (number of elements)
of the first homology group. Examples of invariants of links in 3-manifolds are
given. They exhibit interesting arithmetic properties.

\end{titlepage}

\section{Introduction}

The first purpose of this paper is to carry on the study of the Alexander
Conway
invariant ($\Delta$) from the quantum field theory point of view that was
started in \cite{RS91}. The  goal here, since the Alexander invariant is
well known from the topology point of view, is to
establish a dictionary between the topology and QFT results in that case, and
then to proceed backwards to gain deeper understanding of say the Jones
polynomial or the related 3-manifold invariants that are merely understood
 from the QFT
(or combinatorial) point of view so far. The approach to that problem should a
priori be straightforward. Since after some time the Alexander polynomial has
been put in the right combinatorial framework by introduction of the
quantum super group $U_{q}gl(1,1)$ \cite{KS91,D89,M90}, one
simply has, following the seminal work \cite{W89}, to consider the WZW model
 on the
supergroup $U(1,1)$ and the corresponding Chern  Simons theory.
Difficulties are however met. Representation theory of superalgebras is still
poorly understood in general. Despite various attempts
\cite{GOW87,G89,H90,JZ89,B90,HO90,BMRS91}, the case of affine superalgebras is
also confusing, no good concept of integrability for instance having yet been
exhibited. Moreover the $gl(1,1)$ case seems to present additional difficulties
due to its balance between bosons and fermions (this balance
is deeply related
to the nature of the Alexander invariant\cite{KS91,Lee90}). So far studies of
 WZW models on
supergroups \cite{BMRS91} had focused on (hoped to be  consistent) truncated
theories that involve

only "maximally
atypical" representations. In our case
however such a truncation would provide trivial invariants. The closer study of
$U(1,1)$ WZW model we carry out in this paper shows that, in this problem at
 least,
there is no possible truncation. {\bf All} representations, including typical
 and
atypical representations as well as
indecomposable blocks, have to be considered, and infinite $S$ matrix
 elements send
atypical to blocks. Infinities can be regularized so that, for the level $k$ an
integer, a finite dimensional representation of the modular group is obtained.
Multivariable Alexander Conway invariants can then be computed by surgery, with
results agreeing with classical knot theory literature \cite{K83,BZ85}.
 Moreover invariants of links in 3-manifolds can also be computed.
Casson
invariants \cite{J} based on super Chern Simons theories have  been considered
 earlier
 in \cite{Witten89}.

Besides, the knowledge of applications to topology
 proves to be of great help in the study of WZW models on
supergroups. The second purpose of this paper is to exhibit unexpected features
of such theories  which we believe are generic and not peculiarities
of $U(1,1)$.

The paper is organized as follows. In section 2 we recall some results of
\cite{RS91}, give some formulas for characters and derive part of the $S$
matrix. The complete $S$ and $T$ matrices are derived in section 3 after
consideration of the $U(1,1)$ WZW model on the torus. Sections 2 and 3 are
oriented towards conformal field theory and may be skipped by the reader
interested in applications to topology only. Basic properties of
$S$ and $T$ matrices are discussed in section 4. Surgery is then used in
 section
5 to compute a variety of link invariants, including cable and torus links.
Section 6 is devoted to some (preliminary) study of invariants of links in
3-manifolds and of invariants of 3-manifolds. We also present a discussion of
some results from a three dimensional point of view, making partial contact
with $U(1)$ Casson invariant and the results of \cite{Witten89}.
Some conclusions are gathered in Section 7.

The first appendix makes some partial connection between our results
and the combinatorial approach to 3-manifold invariants. The second
appendix explores further relations between the Alexander Conway polynomial,
Burau representation, and screening contours in the WZW model.

\section{$gl(1,1)^{(1)}$ Representations and characters}

\subsection{Reminder}

This subsection recalls some of the results of \cite{RS91}.

The algebra $gl(1,1)$ has two bosonic and two fermionic
generators obeying the relations
\begin{equation}
\{\psi,\psi^{+}\}=E,\psi^{2}=\left(\psi^{+}\right)^{2}=0,[N,\psi^{+}]=\psi^{+},
[N,\psi]=-\psi
\end{equation}
$E$ being central.
Its irreducible representations have dimension one and two. In the one
dimensional case they are parametrized by a single complex number $n$,
eigenvalue of the $N$ generator. By convention we chose their unique state
to be bosonic, and denote these representations by $(n)$ (in general
superalgebra terminology \cite{Kac1} these are the "atypical").
 In the two dimensional case they are parametrized by a pair of
complex numbers $e(\neq 0),n$ and one has
\begin{equation}
E=\left(\begin{array}{cc}
e&0\\
0&e
\end{array}\right),N=\left(\begin{array}{cc}
n-1&0\\
0&n
\end{array}\right),\psi=\left(\begin{array}{cc}
0&1\\
0&0
\end{array}\right),\psi^{+}=\left(\begin{array}{cc}
0&0\\
e&0
\end{array}\right)
\end{equation}
By convention we chose the state annihilated by $\psi$ to be bosonic, and
denote these representations by $(en)$ (
"typical").  All other representations are four
dimensional indecomposable ones ("blocks") with the schematic structure
 indicated in
figure 1 (they are denoted by $(\widehat{n})$).  The limit $e\rightarrow 0$ of
$(en)$ produces an indecomposable representation ("pseudotypical").
Notice the tensor product
\begin{equation}
(e_{1},n_{1})\otimes(e_{2},n_{2})=(e_{1}+e_{2},n_{1}+n_{2}-1)\ominus
(e_{1}+e_{2},n_{1}+n_{2})
\end{equation}
provided $e_{1}+e_{2}\neq 0$. In the study of the tensor product the second
representation in the right hand side appears with the state annihilated by
$\psi$ being a fermion. We trade it for an equivalent representation where
this state is a boson. This procedure may introduce some minus signs due to
statistics effects, which is indicated by the $\ominus$ symbol. Four
dimensional representations are obtained by
\begin{equation}
(e,n_{1})\otimes(-e,n_{2})=(\widehat{n}),\ n=n_{1}+n_{2}-1
\end{equation}
Notice the vanishing of the super dimension of these two (and four) dimensional
representations.

The $gl(1,1)^{(1)}$ current algebra is defined in the usual way, and
arises from quantization of the $U(1,1)$ W{\cal Z}W model (the
 level $k$ is not
quantized, while $U(1,1)$ can be considered as compact)\footnote{This compacity
should not hide that the metric is indefinite and the corresponding CFT non
unitary, with in fact a spectrum not bounded from below. The $GL(1,1)$ case
behaves identically.}. A noticeable fact is that
$gl(1,1)$ has two casimirs, but a single combination of these emerges after
conformal invariance requirment. The usual shift $k\rightarrow k+c_{v}$ is
replaced by $1/k^{2}$ contributions to the stress energy
tensor. For primary fields associated with one dimensional representations the
conformal weight $h$ vanishes. For primary fields associated with two
dimensional representations one has
\begin{equation}
h_{en}=(2n-1)\frac{e}{2k}+\frac{e^{2}}{2k^{2}}\label{eq:h}
\end{equation}
In four
dimensional indecomposable representations, the casimir is not diagonalizable.
Therefore they have no associated primary field. The value of the conformal
weight, defined eg by considering the trace of the casimir, vanishes.

The current algebra has a free field representation involving a pair of bosonic
fields $\phi_{E},\phi_{\tilde{N}}$ with
\begin{equation}
\left<\phi_{E}(z)\phi_{\tilde{N}}(w)\right>=\left<\phi_{\tilde{N}}(z)\phi_{E}
(w)\right>=
-\mbox{ln}(z-w)
\end{equation}
and a pair of fermionic fields $\eta,\xi$ with
\begin{equation}
\left<\xi(z)\eta(w)\right>=\left<\eta(z)\xi(w)\right>=\frac{1}{z-w}
\end{equation}
In particular the stress energy tensor reads
\begin{equation}
L=-:\eta\partial\xi:-:\partial\phi_{E}\partial\phi_{\tilde{N}}:+\frac{i}{2
\sqrt{k}}\partial^{2}\phi_{E}
\end{equation}
and the central charge is $c=0$. The primary fields $\Phi_{en}$ associated
 with $(en)$
representations are vertex operators
\begin{equation}
V_{en}=\mbox{exp}\frac{i}{\sqrt{k}}\left(\tilde{n}\phi_{E}
+e\phi_{\tilde{N}}\right)
\end{equation}
Where we introduced for further convenience the notation
\begin{equation}
\tilde{n}=n+\frac{e}{2k}
\end{equation}
(as often in this work the notations do not contain explicit reference to all
the parameters involved, for the sake of simplicity. We hope this will not
cause any confusion).
There does not seem to be any simple free field representation associated with
the block representations. The screening operator is
\begin{equation}
{\cal J}=\eta\mbox{ exp}-\frac{i}{\sqrt{k}}\phi_{E}\label{eq:screening}
\end{equation}
As was demonstrated in \cite{RS91}, there is no truncation of the operator
algebra in the W{\cal Z}W model. If for the defining representation
 $e=1,n=1/2$, one
therefore has to deal with all values of $e$ integer and $n$ half integer
for the two and four dimensional
representations, and all values of $n$ half integer
 for the one dimensional ones\footnote{We
do not restrict to the natural parity splitting for later notational
convenience}. Rational values of $k$ are special because then
some of the  conformal weights (\ref{eq:h}) differ by integers and the
$gl(1,1)^{(1)}$ representations can
be regrouped in larger sets, maybe representations themshelves of some bigger
algebra. As we shall see these "extended representations" of $gl(1,1)^{(1)}$
 will provide a finite represenation of the
modular group, and hence allow surgery computations of Alexander invariants for
links and 3-manifolds. In the following we discuss the case
 $k$ {\bf half an odd
integer} only. Introduce for a while
the  notation
\begin{equation}
\lambda_{en}=2n-1+\frac{e}{k},\ \mu_{en}=\frac{e}{2k}
\end{equation}
so that
\begin{equation}
h=\lambda\mu
\end{equation}
Then
\begin{equation}
h=h'\mbox{ mod integer if } \lambda=\lambda'+M,\mu=\mu'+N,
M,N\mbox{ integer}, M+2N=0\mbox{ mod }2k\label{eq:mod}
\end{equation}
Conversely we chose (\ref{eq:mod}) to define which conformal weights have to
be regrouped into extended representations, such that a finite dimensional
 representation
of the modular group will be obtained. The fundamental domain has volume
\begin{equation}
V^{-2}=4k^{2}
\end{equation}
It can be described by $e=0,1,\ldots,2k-1$, $n=1/2,1,3/2,\ldots,k$.

For the $gl(1,1)^{(1)}$
representations associated with the two
and four dimensional representations of $gl(1,1)$, the ordinary supercharacter
defined by $\mbox{Str}(\mbox{exp}2i\pi\tau L_{0})$ vanishes due to boson
fermion symmetry. We are however
interested in the evaluation of the $S$ matrix, which  can be
computed by considering any specialization of the charaters. In the following
we  use the free field representation and obtain non vanishing results by
introducing a background gauge field, or more simply by
suppression of the zero mode of the $\eta\xi$ system.

We introduce
\begin{equation}
\zeta_{en}=\zeta_{\eta\xi}\zeta_{\phi_{E}\phi_{\tilde{N}}}\mbox{exp}
[2i\pi\tau e(\widetilde{n}-1/2)/k]
\end{equation}
In this formula we can consider $\zeta_{\eta\xi}$ to be  the partition
 function of the $\eta\xi$ system
with periodic boundary conditions, the zero mode being subtracted
\begin{equation}
\zeta_{\eta\xi}=\mbox{Im}\tau \eta^{2}\overline{\eta}^{2}
\end{equation}
where $\eta$ is the usual Dedekind function.
It is by itself a modular invariant, and cannot be factorized into the usual
chiral antichiral form due to the $\mbox{Im}\tau$ term. Such factorization is
easy to obtain by introducing a background gauge field ${\cal E}$, which
produces
\begin{equation}
\zeta_{\eta\xi}=\frac{\theta_{1}({\cal E})}{\eta}
\overline{\frac{\theta_{1}({\cal E})}{\eta}}
\end{equation}
Similarly
\begin{equation}
\zeta_{\phi_{E}\phi_{\tilde{N}}}=\frac{1}{\eta^{2}}\label{eq:zboson}
\end{equation}
We set also, from figure 1
\begin{equation}
\zeta_{\widehat{n}}=\zeta_{0,n}-\zeta_{0,n+1}\label{eq:hat}
\end{equation}
To deal now with  representations of $gl(1,1)^{(1)}$ associated with
one dimensional representations of $gl(1,1)$, we consider the latter
as infinite sums of two dimensional ones as in
figure 2. We therefore have to take alternate sum of characters, and also to
take into
account the additional signs coming from statistics. The net result is
\begin{equation}
\zeta_{n}=\sum_{j=0}^{\infty}\zeta_{0,n+1+j}=-\sum_{j=0}^{\infty}
\zeta_{0,n-j}\label{eq:prime}
\end{equation}
This sum is regularized by introduction of the background gauge field ${\cal
E}$. A similar formula appears in \cite{leites}.
For $k$ generic we expect the above to be characters of irreducible
representations of $gl(1,1)^{(1)}$. When $k$ is half an odd integer, the
 screening
operator ${\cal J}$ becomes local with respect to vertex operators $V_{en}$ for
$e=0\mbox{ mod 2k}$. If $e$ does not satisfy this condition, since we have only
one screening operator, which moreover is nilpotent, we expect the above
formulas to be still characters of irreducible $gl(1,1)^{(1)}$ representations
\cite{F89}. If $e=0\mbox{ mod 2k}$ the situation is very similar to what
happens
for $e=0$. Consider for instance $e=-2pk$. Then one finds, using
$J^{\psi}=i\sqrt{k}\xi\partial\phi_{E}+k\partial\xi$
\begin{equation}
J^{\psi}_{-2p}V_{en}=0
\end{equation}
the corresponding state has therefore to be factored out of the free field Fock
space. Its conformal weight is $h_{en}+2p$ which coincides with $h_{e,n-1}$. As
usual the state $V_{e,n-1}$ can be obtained from $V_{en}$ by acting
successively with $J^{\psi}_{0}$ and ${\cal J}$. Considering therefore
$\mbox{Ker}\oint{\cal J}$ in the free field Fock space based on $V_{en}$ should
produce an irreducible representation of $gl(1,1)^{(1)}$. The picture of the
corresponding cohomology is like the second diagram in figure 2 with
 representations $e=-2pk,n$ and the "BRS" operator $\oint{\cal J}$ connecting
$e=-2pk,n$ and $e=-2pk,n-1$. Disappearance of states in the Fock space based on
$V_{e=-2pk,n}$ is associated with appearance of null states in the dual
$V_{e=2pk,1-n}$ (first diagram in figure 2). The characters of irreducible
representations are therefore
\begin{equation}
\zeta_{e=-2pk,n}=-\sum_{j=0}^{\infty}\zeta_{e=-2pk,n-j},\mbox{ and }
\zeta_{e=2pk,n}=\sum_{j=0}^{\infty}\zeta_{e=2pk,n+1+j}
\end{equation}
Both expressions coincide for $p=0$.
For $k$ half an odd integer we define the characters $\chi$ of extended
 representations
of $gl(1,1)^{(1)}$  as
appropriate sums  of the characters $\zeta$ over the two dimensional
lattice (\ref{eq:mod}). Since the spectrum of dimensions is not bounded from
below $\chi$ are naively infinite. They hopefully could be properly defined by
a
treatment analogous to the one of $\beta\gamma$ systems. In the following this
divergence does not apparently cause difficulties. It is not at the
origin
of divergence of some $S$ matrix  elements, as explained later.

\subsection{The naive $S$ matrix}

For $k$ half an odd integer we expect the set of $\chi$'s to provide a finite
dimensional
representation of the modular group. Let us start the computation of the $S$
matrix elements from the above expressions\footnote{The $S$ matrix is defined
here by $\chi_{a}(-1/\tau)=\sum S_{a}^{b}\chi_{b}(\tau)$}. In what follows we
discard the $\eta\xi$ contributions which are easily evaluated and affect the
results by phases only, and we concentrate on the bosonic contributions.
 Using Poisson formula one finds
readily
\begin{equation}
S_{en}^{e'n'}=iV
\mbox{exp}-\frac{2i\pi}{k}\left[e'(\tilde{n}-1/2)+
e(\widetilde{n'}-1/2)\right]\label{eq:S22}
\end{equation}
The usual $\sqrt{\tau}$ term that arises from Poisson formula is compensated by
a similar contribution from the $1/\eta^{2}$ term in (\ref{eq:zboson}).
{}From this, and using the expression (\ref{eq:hat}) one obtains
\begin{equation}
S_{en}^{\widehat{n'}}=-V\mbox{exp}(-2i\pi en'/k)/2\mbox{sin}(\pi
e/k)\label{eq:S2hat}
\end{equation}
while it is likely from boson fermion symmetry that
\begin{equation}
S_{en}^{n'}=0
\end{equation}
(this result could be wrong due to compensation effects. It is fully
established in the next section). Using (\ref{eq:hat}) one can find also
\begin{equation}
S_{\widehat{n}}^{e'n'}=-2V\mbox{sin}(\pi e'/k)\mbox{exp}(-2i\pi e'n/k)
\end{equation}
and using (\ref{eq:prime})(with an infinitesimal background gauge field that
goes to zero at the end of the computation)
\begin{equation}
S_{n}^{e'n'}=V\mbox{exp}(-2i\pi e'n/k)/2\mbox{sin}(\pi
e'/k)\mbox{ "naive"}\label{eq:naive}
\end{equation}
The label "naive" means this formula will be reconsidered by a proper
treatment of the characters associated with one dimensional representations of
$gl(1,1)$ in the next secion. Notice that this element can be non zero because
the representation of
$gl(1,1)$ with vanishing superdimension appears as an upperindex

\section{The $GL(1,1)$ W{\cal Z}W model on a torus}

To obtain the entire $S$ matrix is a delicate matter; we shall see
in particular that some of its elements are infinite (such a result can be
guessed from eg (\ref{eq:naive}) as $e\rightarrow 0$). In order to control and
sometimes regularize things, it is necessary to work on the
torus \cite{G90}. The manipulations we shall carry out now should appear more
 transparent
from the Chern Simons point of view, see section 4.

 We  consider the situation of figure 3, with
a state $|en>$ propagating in the $\tau$ direction, and insertion of a
representation $|e_{0}n_{0}>$ in $z_{1}$, of its conjugate $|-e_{0},1-n_{0}>$
in $z_{2}$. Two conformal blocks contribute, which
 can be computed by putting the free field representatives
\begin{eqnarray}
V_{e_{0}n_{0}}(z_{1})=\mbox{exp}\frac{i}{\sqrt{k}}(\widetilde{n_{0}}\phi_{E}
+e_{0}\phi_{\tilde{N}})\nonumber\\
\xi V_{-e_{0},1-n_{0}}(z_{2})=\xi\mbox{exp}\frac{i}{\sqrt{k}}[
(1-\widetilde{n_{0}})\phi_{E}-e_{0}\phi_{\tilde{N}}]
\end{eqnarray}
together with the screening operator (\ref{eq:screening}) ${\cal J}$. We
 consider first
 integration
on the contour shown on figure 3. The zero modes of the $\eta\xi$ system are
cancelled by the fermionic operators introduced in this construction, and we
obtain the partition function, taking only the chiral contribution from the
bosonic degrees of freedom
\begin{equation}
\zeta_{en}(z_{1},z_{2};\tau)={\cal N}(en)\int_{C}
\zeta_{en}(z_{1},z_{2},w;\tau)dw=\label{eq:E0}
\end{equation}
\[
{\cal N}(en)\mbox{Im}\tau\overline{\eta}^{2}
\int_{C_{1}}  [\theta_{1}(z_{1}-w)/\theta'_{1}(0)]^{-e_{0}/k}
[\theta_{1}(z_{2}-w)/\theta'_{1}(0)]^{e_{0}/k}
[\theta_{1}(z_{1}-z_{2})/\theta'_{1}(0)]^{-2e_{0}
(\widetilde{n_{0}}-1/2)/k}
\]
\begin{equation}
\mbox{exp}\left[\frac{2i\pi}{k}\tau e(\tilde{n}-1/2)+\frac{2i\pi}{k}
e(\widetilde{n_{0}}z_{1}
+(1-\widetilde{n_{0}})z_{2}-w)+\frac{2i\pi}{k}(\tilde{n}-1/2)e_{0}
(z_{1}-z_{2})\right] dw
\end{equation}
where ${\cal N}$ is a normalization factor to be determined.
For the case  under study we should in principle sum such
expressions over the lattice (\ref{eq:mod}) before performing a modular
transformation. In order to make things a little easier for notations we shall
not do so, and deal for a while with continuous Fourier transforms.
The  results of the correct procedure are similar up to the introduction
 of the discrete lattice
(\ref{eq:mod}) and multiplication of the $S$ matrix elements by a factor 1/2
due
to the fact that $e$ labels take integer values, $n$ labels half integer ones.
Also one must treat a little more carefully  the
one and four dimensional representations, corresponding to $e'=0$ in the
integral. With these precautions
we find
\[
\zeta_{en}(z_{1}/\tau,z_{2}/\tau,w/\tau;-1/\tau)=
\]
\begin{equation}\int de'dn'
\frac{i}{k}\mbox{ exp}-\frac{2i\pi}{k}[e'(\tilde{n}-1/2)+e(\widetilde{n'}-1/2)]
\  \tau^{-2e_{0}(\widetilde{n_{0}}-1/2)/k}
\zeta_{e'n'}(z_{1},z_{2},w;\tau)\label{eq:E1}
\end{equation}
However if on the left hand side the contour of integration for the screening
charge looks like in figure 3 (with $\tau$ there being equal to
$-1/\tau$), then on the right hand side we have integration on the contour
called $C_{1}$ in figure 4, resulting in an integral $I_{1}$. We have to
reexpress it in terms of integration along the contours $C_{2}$, $C_{4}$,
giving rise to integrals $I_{2}$, $I_{4}$. First notice that
\begin{equation}
I_{1}-I_{2}+I_{5}-I_{3}+I_{4}=0
\end{equation}
because the total contour has no singular points inside. Now notice that
\begin{equation}
\zeta_{e'n'}(z_{1},z_{2},w+1;\tau)=\mbox{exp}(-2i\pi
e'/k)\zeta_{e'n'}(z_{1},z_{2},w;\tau)
\end{equation}
Similarly
\begin{equation}
\zeta_{e'n'}(z_{1},z_{2},w+\tau;\tau)=\zeta_{e',n'-1}(z_{1},z_{2},w;\tau)
\end{equation}
or
\begin{equation}
I_{2}(e',n')=I_{4}(e',n'-1)
\end{equation}
Therefore
\begin{equation}
I_{1}=\frac{I_{4}(e',n')-I_{4}(e',n'-1)-I_{5}(e',n'-1)}{\mbox{exp}(-2i\pi
e'/k)-1}\label{eq:E6}
\end{equation}
In the limit $e_{0}\rightarrow 0$, we can neglect $I_{5}$. In that limit indeed
the screening operator does not pick up a phase when it goes around the
$(-e_{0},1-n_{0})$ insertion, and the contributions
of the down left and up right parts of
$C_{5}$ cancel each other. Therefore
\begin{equation}
I_{1}=\frac{I_{4}(e'n')-I_{4}(e',n'-1)}{\mbox{exp}(2i\pi e'/k)-1}
,\ e_{0}\rightarrow
0
\end{equation}
and we can rewrite (\ref{eq:E1})
\[
{\cal N}(en)\int_{C} \zeta_{en}(-w/\tau;-1/\tau)dw=
\]
\[
\int de'dn'\  {\cal
N}(en)\ \frac{i}{k}\ \frac{\mbox{exp}-\frac{2i\pi}{k}[e'(\tilde{n}-1/2)+e
(\widetilde{n'}-1/2)]}{\mbox{exp}(-2i\pi
e'/k)-1}\int_{C_{4}}  \left[\zeta_{e'n'}(w;\tau)-\zeta_{e',n'-1}(w;\tau)
\right]dw=
\]
\[
\int
de'dn'\ \frac{i}{k}\ \mbox{exp}-\frac{2i\pi}{k}[e'(\tilde{n}-1/2)
+e(\widetilde{n'}-1/2)]\ \frac{\mbox{exp}(-2i\pi e/k)-1}{\mbox{exp}(-2i\pi
 e'/k)-1}
\]
\begin{equation}
\frac{{\cal N}(en)}{{\cal N}(e'n')} {\cal N}(e'n')\int_{C_{4}}
 \zeta_{e'n'}(w;\tau)dw
\end{equation}
Therefore if we choose
\begin{equation}
{\cal N}(en)=\frac{1}{\mbox{exp}(-2i\pi e/k)-1}\label{eq:E7}
\end{equation}
we obtain
\begin{equation}
\zeta_{en}(-1/\tau)=\int de'dn'\  \frac{i}{k}\ \mbox{exp
}-\frac{2i\pi}{k}[e'(\tilde{n}-1/2)+e(\widetilde{n'}-1/2)]\  \zeta_{e'n'}(\tau)
\end{equation}
and we recover the $S$ matrix element (\ref{eq:S22}). The fact that
 $e_{0}\rightarrow
0$ limit has been taken is implicit in the above notation since the dependence
on $z_{1},z_{2}$ has been suppressed. Also notice that the continuum treatment
does not handle the case $e'=0$ completely. In addition to $\zeta_{e'n'}$
($e'\neq 0$) the modular transformation maps $\zeta_{en}$ also to one and four
dimensional representations, as already discussed above.

Comparaison with (\ref{eq:S2hat}) shows now that
\begin{equation}
\zeta_{\widehat{n'}}(z_{1},z_{2};\tau)=\int_{C_{1}}
\zeta_{0,n'+1}(z_{1},z_{2},w;\tau)dw\label{eq:E2}
\end{equation}

We now argue that
\begin{equation}
\zeta_{n'}(z_{1},z_{2};\tau)\propto\int_{C_{4}}
 \zeta_{0,n'+1}(z_{1},z_{2},w;\tau)
\end{equation}
up to a normalization factor that we shall determine. First we notice that the
integrand is periodic in $w\rightarrow w+1$  if $w$ is not moved between the
insertion points of $(e_{0}n_{0})$ and $(-e_{0},1-n_{0})$ belonging to the
 same torus.
Therefore we can move $C_{4}^{0}$ up to $C_{6}^{1}$ (see figure 5) and
\begin{equation}
\int_{C_{4}^{0}}\zeta_{0,n'+1}(z_{1},z_{2},w;\tau)dw=\int_{C_{6}^{1}}
\zeta_{0,n'+1}(z_{1},z_{2},w;\tau)dw=\int_{C_{6}^{0}}
\zeta_{0n'}(z_{1},z_{2},w;\tau)dw
\end{equation}
Now using the fact that
\begin{equation}
-\int_{C_{4}^{i}}\zeta_{0n'}(z_{1},z_{2},w;\tau)dw+\int_{C_{6}^{i+1}}
\zeta_{0n'}(z_{1},z_{2},w;\tau)dw=0
\end{equation}
we can add and subtract these integrals to get
\begin{equation}
\int_{C_{4}^{0}}\zeta_{0,n'+1}(z_{1},z_{2},w;\tau)dw=\sum_{j=0}^{\infty}
\left(\int_{C_{6}^{0}}
-\int_{C_{4}^{0}}\right)\zeta_{0,n'-j}(z_{1},z_{2},w;\tau)dw
\end{equation}
Now (figure 6)
\begin{equation}
\left(\int_{C_{6}}-\int_{C_{4}}\right)\zeta_{0,n'-j}(z_{1},z_{2},w;\tau)dw=
\left(
\int_{C_{7}}-\int_{C_{8}}\right)\zeta_{0,n'-j}(z_{1},z_{2},w;\tau)dw
\end{equation}
However
\begin{equation}
\zeta_{0,n'-j}(z_{1},z_{2},w-1;\tau)=\mbox{exp}(-2i\pi
e_{0}/k)\zeta_{0,n'-j}(z_{1},z_{2},w;\tau)
\end{equation}
because in that case the point $w$ is moved between the insertions of
$(e_{0}n_{0})$ and $(-e_{0},1-n_{0})$. Therefore
\begin{equation}
\int_{C_{8}}\zeta_{0,n'-j}(z_{1},z_{2},w;\tau)dw=\mbox{exp}(-2i\pi e_{0}/k)
\int_{C_{7}}\zeta_{0,n'-j}(z_{1},z_{2},w;\tau)dw
\end{equation}
and we can write
\begin{equation}
\int_{C_{4}}\zeta_{0,n'+1}(z_{1},z_{2},w;\tau)dw=\left[1-\mbox{exp}(-2i\pi
e_{0}/k)\right]\sum_{j=0}^{\infty}\int_{C_{7}}\zeta_{0,n'-j}
(z_{1},z_{2},w;\tau)dw
\end{equation}
We now set
\begin{eqnarray}
\zeta_{n'}(z_{1},z_{2};\tau)&=&-\mbox{exp}(-i\pi e_{0}/k)
\sum_{j=0}^{\infty}\int_{C_{7}}\zeta_{0,n'-j}
(z_{1},z_{2},w;\tau)dw\label{eq:E3}\\
&=&-\frac{\mbox{exp}(-i\pi e_{0}/k)}{1-\mbox{exp}(-2i\pi
e_{0}/k)}\int_{C_{4}}\zeta_{0,n'+1}(z_{1},z_{2},w;\tau)dw\label{eq:E4}
\end{eqnarray}
Formulas (\ref{eq:E2}) and (\ref{eq:E3},\ref{eq:E4}) provide explicit
 expressions for the
characters of  four and one dimensional representations.
 We can now carry out the modular transformation
\begin{eqnarray}
\zeta_{n}(z_{1}/\tau,z_{2}/\tau;-1/\tau)&=&-\frac{\mbox{exp}(-i\pi e_{0}/k)}
{1-\mbox{exp}(-2i\pi
e_{0}/k)}\tau^{-2e_{0}(\widetilde{n_{0}}-1/2)}\ \frac{i}{k} \int dn'
 \int_{C_{1}}\zeta_{0n'}(z_{1},z_{2},w;\tau)dw\nonumber\\
&+&\mbox{ other terms}\label{eq:E5}
\end{eqnarray}
(the other terms will be analyzed later on) ie from (\ref{eq:E2})
\begin{eqnarray}
\zeta_{n}(z_{1}/\tau,z_{2}/\tau;-1/\tau)&=&-\frac{\mbox{exp}(-i\pi e_{0}/k)}
{1-\mbox{exp}(-2i\pi
e_{0}/k)}\tau^{-2e_{0}(\widetilde{n_{0}}-1/2)}\ \frac{i}{k}
\int dn' \zeta_{\widehat{n'}}(z_{1},z_{2},\tau)\nonumber\\
&+&\mbox{ other terms}
\end{eqnarray}
If we take the limit $e_{0}\rightarrow 0$ in that expression, the
prefactore diverges, meaning that the corresponding $S$ matrix element
diverges (the origin of this divergence lies in the fact that atypical
representations must be considered as infinite sums of typical ones).
We shall write symbolically
\begin{equation}
S_{n}^{\widehat{n'}}=-\frac{iV\mbox{exp}(-i\pi e_{0}/k)}{1-\mbox{exp}(-2i\pi
e_{0}/k)}=-\frac{V}{2\mbox{sin}(\pi e_{0}/k)}=-\frac{V}{2\epsilon}
\end{equation}
The meaning of the insertions necessary to regularize this expression will be
explained in the discussion of link invariants later on. In a similar fashion
we find
\begin{eqnarray}
\zeta_{\hat{n}}(z_{1}/\tau,z_{2}/\tau;-1/\tau)&=&\mbox{exp}(i\pi e_{0}/k)
[1-\mbox{exp}(-2i\pi
e_{0}/k)]\tau^{-2e_{0}(\widetilde{n_{0}}-1/2)}\frac{i}{k}\  \int dn'
 \zeta_{n'}(z_{1},z_{2};\tau)\nonumber\\
&+&\mbox{ other terms}
\end{eqnarray}
and therefore
\begin{equation}
S_{\widehat{n}}^{n'}=-2V\mbox{sin}(\pi e_{0}/k)=-2V\epsilon
\end{equation}
In the following we shall refer to the character (\ref{eq:E0}) as
 $\zeta_{en/I}$,
because  we need to introduce still another specie of character. It is
obtained by a definition similar to (\ref{eq:E0}), but without normalization
factors, and with the contour $C_{7}$ of figure 6, corresponding to the other
available conformal block.
We refer to this new specie as
${\cal Z}_{en/II}$. The fact that there are two conformal blocks on the torus
 for
insertion of $(e_{0}n_{0})$, $(-e_{0},1-n_{0})$ is because the product of these
two representations, although giving only $(\widehat{0})$ at the classical
level of $gl(1,1)$ tensor product, produces also $(0)$ representation at the
quantum level (see the study of four point functions in \cite{RS91}).
The contour $C_{7}$ is unchanged in modular transformation, and we
therefore have immediately
\begin{equation}
S_{en/II}^{e'n'/II}=S_{en/I}^{e'n'/I}=iV\mbox{exp}-\frac{2i\pi}{k}[e'
(\tilde{n}-1/2)+e(\widetilde{n'}-1/2)]
\end{equation}
The reason for introducing such characters is the study of the remaining
 terms in
(\ref{eq:E5}). It is difficult to use the form (\ref{eq:E4})
 because of the vanishing
normalization factor ${\cal N}(0n')$. Let us therefore consider the form
(\ref{eq:E3}). By invariance of the contour $C_{7}$ we find
\[
\zeta_{n}(-z_{1}/\tau,-z_{2}/\tau;-1/\tau)
\]
\begin{equation}
=-\sum_{j=0}^{\infty}\int de'dn'
\frac{i}{k}\mbox{exp}-[2i\pi e'(n-j-1/2)/k]\int_{C_{7}}
 \zeta_{e'n'}(z_{1},z_{2},w;\tau)dw
\end{equation}
ie
\begin{equation}
S_{n}^{e'n'/I}=0
\end{equation}
and
\begin{equation}
S_{n}^{e'n'/II}=V\mbox{exp}(-2i\pi e'n/k)/2\mbox{sin}(\pi e'/k)
\end{equation}
By similar methods one finds
\begin{equation}
S_{en/II}^{\widehat{n'}}=0
\end{equation}
and
\begin{equation}
S_{en/II}^{n'}=2V\mbox{sin}(\pi e/k)\mbox{exp}(2i\pi en'/k)
\end{equation}
and
\begin{equation}
S_{\widehat{n}}^{e'n'/II}=0
\end{equation}
Finally one has also
\begin{equation}
S_{\widehat{n}}^{\widehat{n'}}=0,S_{n}^{n'}=0
\end{equation}
Notice that the fusion of type $II$ representations does not lead to new block
 representations, which is why we restrict to one specie of
$(\widehat{n})$ representations \footnote{One has
$(en/II).(-en'/II)=-2(n+n'-1)+(n+n')+(n+n'-2)$}.
These $S$ matrix elements are invariant under the translations (\ref{eq:mod}).
The correct procedure using discrete sums over the lattice (\ref{eq:mod}) shows
that they are the $S$ matrix elements for the
characters of extended representations $\chi$.
For convenience we summarize our results  in the following
table
\[
S_{en/I}^{e'n'/I}=iV\mbox{exp}-\frac{2i\pi}{k}[e'(\widetilde{n}-1/2)
+e(\widetilde{n'}-1/2)],\ S_{en/I}^{e'n'/II}=0
\]
\[
S_{en/I}^{\widehat{n'}}=-V\mbox{exp}(-2i\pi en'/k)/2\mbox{sin}(\pi e/k),\
S_{en/I}^{n'}=0
\]
\[
S_{en/II}^{e'n'/II}=iV\mbox{exp}-\frac{2i\pi}{k}[e'(\widetilde{n}-1/2)
+e(\widetilde{n'}-1/2)],\ S_{en/II}^{e'n'/I}=0
\]
\[
S_{en/II}^{\widehat{n'}}=0,\ S_{en/II}^{n'}=2V\mbox{sin}(\pi
e/k)\mbox{exp}(-2i\pi en'/k)
\]
\[
S_{\widehat{n}}^{e'n'/I}=-2V\mbox{sin}(\pi e'/k)\mbox{exp}(-2i\pi e'n/k),\
S_{\widehat{n}}^{e'n'/II}=0
\]
\[
S_{\widehat{n}}^{\widehat{n'}}=0,\ S_{\widehat{n}}^{n'}=0\ (reg.:-2V\epsilon)
\]
\[
S_{n}^{e'n'/I}=0,\ S_{n}^{e'n'/II}=V\mbox{exp}(-2i\pi e'n/k)/2\mbox{sin}(\pi
e'/k)
\]
\begin{equation}
S_{n}^{\widehat{n'}}=\infty\ (reg.:-V/2\epsilon),
\ S_{n}^{n'}=0\label{eq:Smatrix}
\end{equation}
where we have set
\begin{equation}
\mbox{sin}\pi e_{0}/k=\epsilon
\end{equation}
For $S$ matrix elements which remain finite as $e_{0}\rightarrow 0$, we have
taken this limit. For an ordinary W{\cal Z}W model, the partition functions
corresponding to $\zeta_{en/I}$ and $\zeta_{en/II}$ would both tend
towards the same character $\zeta_{en}$ as $e_{0}\rightarrow 0$, and the
 effect of the
insertions disappear. In our case such a limit cannot be taken because
divergences occur. To control them we need to keep a small but non zero value
of $e_{0}$. Hence strictly speaking we are not dealing with characters but
conformal blocks, and this implies a doubling of the species. This
 regularization has a simple geometrical meaning from the point of view of
Chern Simons theories which we shall consider in section 5.

Let us now discuss a bit more the $(\widehat{n})$ representations. Using
 the  normalization (\ref{eq:E7}) we can study the
 $e'\rightarrow 0$
limit of  (\ref{eq:E6}) which becomes
\begin{equation}
I_{1}=-\frac{k}{2i\pi}\left[\partial_{e'}I_{4}(0,n')-\partial_{e'}I_{4}(0,n'-1)-
\partial_{e'}I_{5}(0,n'-1)\right]
\end{equation}
while from the definitions
\[
\frac{k}{2i\pi}\partial_{e'}\zeta_{0n'}(z_{1},z_{2},w,\tau)=
\]
\begin{equation}
\tau
(\widetilde{n'}-1/2)\zeta_{0n'}(z_{1},z_{2},w;\tau)+[\widetilde{n_{0}}z_{1}
+(1-\widetilde{n_{0}}z_{2}]\zeta_{0n'}(z_{1},z_{2},w;\tau)-w\zeta_{0n'}
(z_{1},z_{2},w;\tau)\label{eq:E8}
\end{equation}
Now recall that $\zeta_{0n}(z_{1},z_{2},w;\tau)$ is periodic under
$w\rightarrow
w+1$ provided $w$ is not moved between $z_{1}$ and $z_{2}$. Therefore the
combination of integrations
\[
\int_{C_{4}}-\int_{C_{2}}-\int_{C_{5}}
\]
leads to a vanishing contribution of the second term in the right hand side of
(\ref{eq:E8}). As for the first term its contribution to $I_{1}$
is easily evaluated to be
\begin{equation}
\tau \int_{C_{4}}\zeta_{0n'}(z_{1},z_{2},w;\tau)dw\label{eq:E9}
\end{equation}
and due to (\ref{eq:E4}) it vanishes as $e_{0}\rightarrow 0$. Therefore the
main
contribution comes from the last term in (\ref{eq:E8}). Nevertheless
(\ref{eq:E9}) is important because it indicates a non naive behaviour of
$\zeta_{\widehat{n'}}$ in the transformation $\tau\rightarrow\tau +1$. We
find \footnote{The $T$ matrix is defined by the conventions
$\chi_{a}(\tau+1)=\sum T_{a}^{b}\chi_{b}(\tau)$}
\begin{equation}
T_{\widehat{n}}^{n'}=-2iV\ \mbox{sin}(\pi e_{0}/k)\mbox{exp}(-i\pi/6)
\end{equation}
We write next the complete $T$ matrix (recall we have discarded the $\eta\xi$
contributions so the apparent central charge is $c=2$)
\[
T_{en/I}^{en/I}=\mbox{exp}(-i\pi/6)\mbox{exp}[2i\pi e(\widetilde{n}-1/2)/k],\
T_{en/I}^{\mbox{others}}=0
\]
\[
T_{en/II}^{en/II}=\mbox{exp}(-i\pi/6)\mbox{exp}[2i\pi e(\widetilde{n}-1/2)/k],\
T_{en/II}^{\mbox{others}}=0
\]
\[
T_{\widehat{n}}^{\widehat{n}}=\mbox{exp}(-i\pi/6),\
T_{\widehat{n}}^{n'}=0(reg.:-2i\epsilon V\mbox{exp}(-i\pi/6))
,\ T_{\widehat{n}}^{\mbox{others}}=0
\]
\begin{equation}
T_{n}^{n}=\mbox{exp}(-i\pi/6),
\ T_{n}^{\mbox{others}}=0\label{eq:Tmatrix}
\end{equation}

Instead of remembering that the above results apply to conformal blocks with
insertion of operators with small charge $\pm e_{0}$, a more axiomatic,
 easier to use, point of
view is to consider the partition functions as genuine characters. One then
 needs
to assume
 that quantization has led to a doubling of species: while in $gl(1,1)$ $(en)$
and $(\widehat{n})$ representations form a closed set in tensor product,
quantization introduces also $(n)$ and the splitting $(en)\rightarrow
(en/I), (en/II)$. Then one takes the limit $e_{0}\rightarrow 0$ so that some
$S$ matrix elements vanish, others become infinite, and the presence of
$\epsilon$
in the above formulas is merely giving a rule for multiplying $0\times\infty$.
To check whether this is consistent we now discuss in details the properties of
this $S$ matrix.

\section{Properties of the $S$ matrix}

\subsection{$S^{2}=C$}

Let us now consider the square of the $S$ matrix. Start with a representation
$(en/I)$ and apply $S$ twice. We have various possible sequences
\begin{eqnarray*}
(i)\ en/I\rightarrow e'n'/I\rightarrow e''n''/I\\
(ii)\ en/I\rightarrow e'n'/I\rightarrow \widehat{n''}\\
(iii)\ en/I\rightarrow \widehat{n'}\rightarrow e''n''/I\\
(iiii)\ en/I\rightarrow \widehat{n'}\rightarrow n''
\end{eqnarray*}
Sequences $(i)$ and $(iii)$ contribute as
\begin{eqnarray}
\left(S^{2}\right)_{en/I}^{e''n''/I}&=&-V^{2}\sum^{*}_{e'n'}
\mbox{exp}-\frac{2i\pi}{k}
[e'(\widetilde{n}+\widetilde{n''}-1)+(e+e'')(\widetilde{n'}-1/2)]\nonumber\\
&+&V^{2}\sum_{n'}\frac{\mbox{sin}(\pi e''/k)}{\mbox{sin}(\pi
e/k)}\mbox{exp}-\frac{2i\pi}{k}n'(e+e'')
\end{eqnarray}
where the $*$ sum means the value $e'=0$ is excluded (and of course $e,e''$ are
non zero). Because the sum over $n'$ is unrestricted it imposes $e+e''=0$, and
the second sum then contributes just what is missed in the first sum by the $*$
restriction to provide a function $-\delta_{n+n''-1}$ (modulo the lattice
(\ref{eq:mod})(the dependence of
$\widetilde{n}$ upon $e$ does not influence the results of Fourier sums which
can be as well evaluated as if $\widetilde{n}$ was an independent variable).
 The sequence $(ii)$ does
not contribute since there the sum over $n'$ imposes $e=0$, which is excluded.
The sequence $(iiii)$ vanishes as $\epsilon\rightarrow 0$. Therefore
\begin{equation}
\left(S^{2}\right)_{en/I}^{-e,1-n/I}=-1,\ 0\mbox{ otherwise}
\end{equation}
The minus sign occurs for the reason that in the tensor product of
representations $(en)$ and $(-e,1-n)$ both with the upper state being a boson,
the $gl(1,1)$ invariant state generated is a fermion. In our conventions where
one dimensional representations are made of a bosonic state, we trade it for a
boson, and need therefore a minus sign in the definition of charge
conjugation.
Starting with a  representation $(en/II)$  we find similarly
\begin{equation}
\left(S^{2}\right)_{en/II}^{-e,1-n/II}=-1,\ 0\mbox{ otherwise}
\end{equation}
Consider now the action of $S^{2}$ on a representation $\widehat{n}$. Again
four possible sequences can occur
\begin{eqnarray*}
(i)\ \widehat{n}\rightarrow e'n'/I\rightarrow \widehat{n''}\\
(ii)\ \widehat{n}\rightarrow e'n'/I\rightarrow e''n''/I\\
(iii)\ \widehat{n}\rightarrow n'\rightarrow \widehat{n''}\\
(iiii)\ \widehat{n}\rightarrow n'\rightarrow e''n''/II
\end{eqnarray*}
Sequences $(i)$ and $(iii)$ combine to give
\begin{eqnarray}
\left(S^{2}\right)_{\widehat{n}}^{\widehat{n''}}&=&V^{2}\sum^{*}_{e'n'}
\mbox{exp}-\frac{2i\pi}{k} e'(n+n'')\nonumber\\
&+& V^{2}\sum_{n'} 1
\end{eqnarray}
As before the second term provides the missing $e'=0$ contribution in the first
sum to give $\delta_{n+n''}$. It is therefore essential to consider it to
 insure $S^{2}=C$. Sequence
$(ii)$ does not contribute because the sum over $n'$ imposes $e''=0$, which is
excluded. Sequence $(iiii)$ does not contribute in the limit
 $\epsilon\rightarrow
0$ where the first $S$ matrix element vanishes. Therefore
\begin{equation}
\left(S^{2}\right)_{\widehat{n}}^{\widehat{-n}}=1,\ 0\mbox{ otherwise}
\end{equation}
 The result for
$(n)$ representations is entirely similar
\begin{equation}
\left(S^{2}\right)_{n}^{-n}=1,\ 0\mbox{ otherwise}
\end{equation}
With the above rules to take conjugates of two dimensional representations, one
checks that indeed $(\widehat{-n})$ is the conjugate of $(\widehat{n})$, and
 similarly
$(-n)$ the conjugate of $(n)$, so $S^{2}=C$ is established.
Let us  summarize the action of charge conjugation for completeness
\begin{equation}
(en)\rightarrow\ -(-e,1-n),\
(\widehat{n})\rightarrow (\widehat{-n}),\
(n)\rightarrow (-n)
\end{equation}
In the Chern Simons point of view, taking conjugate of the representation
 carried by a strand is like keeping this representation but changing the
strand orientation. One checks that $SC=CS$ and $SC=S^{*}$ where $*$ means
complex conjugation.

\subsection{ $(ST)^{3}=C$}

The computation of $(ST)^{3}$ can be done using the above formulas. As before
the dependence of $\widetilde{n}$ upon $e$ does not modify naive results
obtained by treating it as an independent variable. Let us discuss for instance
the matrix element between $(en)$ and $(e'n')$. Several sequences can
contribute
\begin{eqnarray*}
(i) en/I\rightarrow e_{1}n_{1}/I\rightarrow e_{2}n_{2}/I\rightarrow e'n'/I\\
(ii) en/I\rightarrow \widehat{n_{1}}\rightarrow e_{2}n_{2}/I\rightarrow
e'n'/I\\
(iii) en/I\rightarrow e_{1}n_{1}/I\rightarrow \widehat{n_{2}}
\rightarrow e'n'/I\\
\end{eqnarray*}
where the arrows represent action with $ST$.
As above the sequences $(ii),(iii)$
complete the missing terms in $(i)$ to restore
the entire sum of the discrete fourier transform, and one finds
\begin{equation}
\left[(ST)^{3}\right]_{en/I}^{-e,1-n/I}=-1,\ 0\mbox{ otherwise}
\end{equation}
An identical result is obtained with the representations of type $II$, the role
of $(\widehat{n})$ representations as intermediates being played by $(n)$
representations. The infinitesimal branching
of $(\widehat{n})$ to $(n')$ in $T$ and infinitely large branching of $(n')$ to
$(\widehat{n''})$ in $S$ produces also sequences involving only
 $(\widehat{n})$ and
$(n)$ representations. For instance
\[
n\rightarrow -\frac{V}{2\epsilon}\widehat{n_{1}}\rightarrow
V^{2}n_{2}-\frac{iV^{2}}{2\epsilon}\widehat{n_{2}}\rightarrow
iV^{3}n'
\]
(we did not write the additional phase factor $\mbox{exp}(-3i\pi/6)$ coming
from
$T$ transformations). Again such sequences just compensate for the missing
terms in the sums over two dimensional representations to reproduce fourier
transforms and one gets in every case
\begin{equation}
(ST)^{3}=C
\end{equation}
Acting on $gl(1,1)^{(1)}$ extended characters,
"doubled" by the above regularization procedure,
we have therefore obtained a finite dimensional representation
of the modular group for $k$ half an odd integer \footnote{This is true in
 the limit
$\epsilon\rightarrow 0$ only}.

In \cite{RS91}, the $S$ matrix elements $S_{en/I}^{en/I}$ had been correctly
obtained using Verlinde formula (we shall discuss this again later). We had not
treated the block representations in an independent way but rather
using the pseudotypical $(0n)$ representations, which thanks to (\ref{eq:hat})
amounts to the same thing. Some of the $S$ matrix elements involving
 representations $(n)$
were only deduced from consistency of Verlinde formula, and the whole $S$
matrix was not obtained. This is now completed, and will allow us later on to
compute links and 3-manifold invariants by surgery.

\subsection{Metric and unitarity of the $S$ matrix}

In the computation of link invariants a crucial role is played by the partition
function of Chern Simons theory on $S^{2}\times S^{1}$ with two Wilson lines
carrying different representations. This defines a metric for $gl(1,1)^{(1)}$
extended representations that we shall now determine (figure 7). We start by
 the convention
\begin{equation}
<en/I|e'n'/I>=\delta_{e+e'}\delta_{n+n'-1}
\end{equation}
$(\widehat{n})$ representations appearing as combinations of $(0n/I)$
representations we can set also
\begin{equation}
<\widehat{n}|\widehat{n'}>=-2\delta_{n+n'}+\delta_{n+n'+1}+
\delta_{n+n'-1}\label{eq:tensor}
\end{equation}
\begin{equation}
<en/I|\widehat{n'}>=0
\end{equation}
Considering $(n)$ as infinite sum of $(0n/I)$ representations we find also
\begin{equation}
<\widehat{n}|n'>=\delta_{n+n'}
\end{equation}
\begin{equation}
<en/I|n'>=0
\end{equation}
We now have to set the following result that looks a priori surprising
\begin{equation}
<n|n'>=0 \label{eq:surprise}
\end{equation}
This implies in particular that for the sphere with two insertions of the
 fields associated with one dimensional representations of $gl(1,1)$
 , the Hilbert space of the Chern Simons theory has vanishing dimension, or
that
there is no conformal block. Such result can be derived from the computation of
the four point functions done in \cite{RS91}. In this reference in particular
\[
<\Phi_{en_{1}}(\infty)\Phi_{-e,n_{2}}(z)\Phi_{e'n_{3}}(0)\Phi_{-e',n_{4}}(1)>
\]
was determined. It involves two conformal blocks called ${\cal F}^{1}$ and
${\cal F}^{2}$ such that only ${\cal F}^{1}-{\cal F}^{2}$ does not have
 logarithm at
infinity. The latter combination is interpreted as representing the
 contribution of a  field
associated with a four  dimensional block representation of $gl(1,1)$
 in the product
$\Phi_{en}(\infty)\Phi_{-e,1-n}(z)$. The other independent block say ${\cal
F}^{1}$ does contain a logarithm at infinity and represents, added generically
to the above, the contribution of
the  field associated with a one dimensional representation of
$gl(1,1)$ in this product. Now the physical correlator was found to be
\[
G\propto ({\cal F}^{1}-{\cal F}^{2})(\overline{{\cal F}^{1}}-\overline{
{\cal F}^{2}})-\left[{\cal F}^{1}(\overline{{\cal F}^{1}}-\overline{
{\cal F}^{2}})+\overline{{\cal F}^{1}}({\cal F}_{1}-{\cal F}_{2})\right]
\]
In this expression the conformal blocks for four dimensional representations
are coupled together and with the conformal blocks of one dimensional
representations. The latter however are not coupled, in agreement with
(\ref{eq:surprise}).

Various manipulations lead to the complete set of scalar products
\[
<en/I|e'n'/I>=\delta_{e+e'}\delta_{n+n'-1}=
<en/I|e'n'/II>,<en/II|e'n'/II>=0
\]
\[
<en/I|n'>=<en/II|\widehat{n'}>=0
\]
\[
<en/II|n'>=<en/II|\widehat{n'}>=0
\]
\[
<n|n'>=0,<n|\widehat{n'}>=\delta_{n+n'}
\]
\begin{equation}
<\widehat{n}|\widehat{n'}>=-2\delta_{n+n'}+\delta_{n+n'+1}+\delta_{n+n'-1}
\end{equation}
This defines a metric $g$ which is real
symmetric. Using it one can in particular
 lower indices. Notice that the metric is not given by the complex
conjugation matrix. We now have to check that the scalar product is invariant
under modular transformations ie
\begin{equation}
<S\rho_{i}|S\rho_{j}>=<\rho_{i}|\rho_{j}>
\end{equation}
Let us discuss a few examples. Consider first
\[
<n|n'>=0
\]
and apply $S$ to both states. Under $S$ a representation $(n)$ branches to
representations $(e_{1}n_{1}/II)$ and $(\widehat{n_{1}})$. Because the first
are
mutually orthogonal, as are the first and the second, we get
\[
<Sn|Sn'>=\sum_{n_{1},n_{2}}<\widehat{n_{1}}|\widehat{n_{2}}>=0
\]
since the trace of (\ref{eq:tensor}) vanishes. Consider now
\[
<n|\widehat{n'}>=\delta_{n+n'}
\]
One finds
\[
<Sn|S\widehat{n'}>=-V^{2}\sum^{*}_{e_{1}n_{1}e_{2}n_{2}}
\mbox{exp}-[2i\pi(-ne_{1}+n'e_{2})/k]
\mbox{sin}(\pi e_{2}/k)/\mbox{sin}(\pi
e_{1}/k)\delta_{e_{1}+e_{2}}\delta_{n_{1}+n_{2}-1}
\]
\[
+V^{2}\sum_{n_{1}n_{2}}\delta_{n_{1}+n_{2}}
\]
as usual the last sum contributes the missing part in the first one to
reproduce $\delta_{n+n'}$ (modulo the lattice (\ref{eq:mod})).
 Other cases are checked in a similar fashion.

\section{Link invariants and the Verlinde formula}

Here as in \cite{RS91} we assume that the relation between Chern
Simons and WZW theories established in the compact, non graded case
 \cite{W89,EL89} extends here.
Call generically ${\cal Z}$ the partition function of the Chern Simons theory
for some link in $S^{3}$. According to \cite{RS91}, introduce
\begin{equation}
\Box'=\prod_{\mbox{crossings}}\mbox{exp}-i\pi\epsilon[e'(\widetilde{n}-1/2)
+e(\widetilde{n'}-1/2]\ {\cal Z}\label{eq:phase}
\end{equation}
where $\epsilon$ is the sign of a crossing where representations $(en)$ and
$(e'n')$ cross (independently of their type)\footnote{(if one strand carries a
 four
dimensional representation, this formula can still be used by considering it as
the concatenation of two two dimensional ones)}
. Then the following holds
\begin{equation}
\Box'=iV\Delta
\end{equation}
where $\Delta$ is the multivariable Alexander Conway polynomial, whose precise
axiomatic definition can be found in \cite{RS91}, with parameters
$t_{i}=q^{e_{i}}, q=\mbox{exp}(-i\pi /k)$\footnote{The reader may notice some
 changes of sign with
respect to \cite{RS91}. This is because in the present paper we use the CFT
conventions for $1,\tau$ instead of the more usual knot theoretic conventions
that invert the torus meridian.}.

\subsection{Hopf links}

As a first example we consider two linked circles, one carrying a
representation $(en/I)$ and the other $(n')$. Suppose we make modular
transformation on the circle carrying $(en/I)$. We find then $S^{2}\times
S^{1}$ with two parallel loops, one carrying the images of $(en/I)$ and the
other $(n')$. Only the $(\widehat{n''})$ representations couple to $(n')$
 among the
images of $(en/I)$. Due to $<n|\widehat{n'}>=\delta_{n+n'}$ one finds
\begin{equation}
{\cal Z}=-V\mbox{exp}(2i\pi en'/k)/2\mbox{sin}(\pi e/k)
\end{equation}
We can also make modular transformation on the circle carrying $(n')$. Only the
$(e''n''/II)$ representations couple to $(en/I)$ among the images of $(n')$.
Due
to $<e''n''/II|en/I>=\delta_{e+e''}\delta_{n+n''-1}$ one finds
\begin{equation}
{\cal Z}=V\mbox{exp}(2i\pi en'/k)/2\mbox{sin}(-\pi e/k)
\end{equation}
coinciding with the above result. This common value is of course equal to
$S_{en/I,n'}$.

As a second example we consider again  two linked circles, one carrying
$(\widehat{n})$ and the other $(n')$. Making modular transformation on the
 circle
carrying $(\widehat{n})$ we find a vanishing result since $(\widehat{n})$
branches only to $(e''n''/I)$ and $(n'')$, both having vanishing scalar product
with $(n')$. Making modular tranformation on the circle that carries $(n')$
, only the branching from $(n')$ to $(\widehat{n''})$ contributes. However
since
the trace of (\ref{eq:tensor}) vanishes, the total evaluation is still zero.
 This
gives the value of $S_{\widehat{n},n'}$.

As a third example consider again two linked circles, one carrying $(en/I)$ and
the other $(\widehat{n'})$. Suppose we make modular transformation on the
circle
carrying $(en/I)$. Only the branching to $(\widehat{n''})$ contributes. One
finds
\[
{\cal Z}=-\frac{V}{2\mbox{sin}\pi e/k}\left\{-2\mbox{exp}(2i\pi en'/k)
+\mbox{exp} [2i\pi
e(n'-1)/k]+\mbox{exp} [2i\pi e(n'+1)/k]\right\}
\]
ie
\begin{equation}
{\cal Z}=2V\mbox{sin}(\pi e/k)\mbox{exp}(2i\pi en'/k)
\end{equation}
If we make modular transformation on $(\widehat{n'})$, only the branching to
$(e''n''/I)$ contributes. One finds
\[
{\cal Z}=2V\mbox{sin}(\pi e/k)\mbox{exp}(2i\pi en'/k)
\]
as above. We notice that such value could as well be recovered by using the
 $U_{q}gl(1,1)$ formalism. Since in  \cite{RS91} the universal $R$ matrix was
obtained, one has simply to evaluate it when acting in the product of an
block representation and a two dimensional one, and take properly
normalized trace. Corresponding to table (\ref{eq:Smatrix})
 we now give the entire set of Hopf
links (figure 8) partition functions
\[
S_{en/I,e'n'/I}=iV\mbox{exp}\frac{2i\pi}{k}\left[e'(\widetilde{n}-1/2)+
e(\widetilde{n'}-1/2)\right]=\ S_{en/I,e'n'/II}
\]
\[
S_{en/I,\widehat{n'}}=2V\mbox{sin}(\pi e/k)\mbox{exp}(2i\pi en'/k),\
S_{en/I,n'}=-V\mbox{exp}(2i\pi en'/k)/2\mbox{sin}(\pi e/k)
\]
\[
S_{en/II,e'n'/II}=0
\]
\[
S_{en/II,\widehat{n'}}=2V\mbox{sin}(\pi e/k)\mbox{exp}(2i\pi en'/k),\
S_{en/II,n'}=0
\]
\[
S_{\widehat{n},\widehat{n'}}=0\ (reg.:-2V\epsilon),\ S_{\widehat{n},n'}=0
\]
\begin{equation}
S_{n,n'}=\infty\ (reg.:-V/2\epsilon)\label{eq:Hopf}
\end{equation}
the others being deduced by symmetry
$S_{\rho_{i}\rho_{j}}=S_{\rho_{j}\rho_{i}}$. From the formal point of view
where the
regularization is forgotten and $\epsilon$ is interpreted as coding the
multiplication rules of $0\times\infty$, we see that some link invariants have
to be set equal to infinity. As was commented in \cite{RS91} this is rather
natural from the point of view of the axiomatic construction of the
multivariable Alexander polynomial (see also the appendix).

\subsection{Geometrical meaning of the regularization}

It is now time to comment more on these results by looking at the geometrical
meaning of the regularization introduced above. The partition function
${\cal Z}_{en/I}(z_{1},z_{2};\tau)$ corresponds to the situation depicted in
 figure
9. We have a solid torus $D^{2}\times S^{1}$ with a Wilson loop carrying the
representation $(en)$. Attached to this main
line by trivalent vertices \cite{W89}
are two dotted lines with opposite orientation carrying $(e_{0}n_{0})$
 which cut the surface
of the torus at points
$z_{1}$ and $z_{2}$. The partition function ${\cal
Z}_{en/II}(z_{1},z_{2};\tau)$
corresponds to the situation depicted in figure 10. The difference is that
there is a single  dotted line carrying $(e_{0}n_{0})$, that intersects the
surface of the torus at $z_{1}$ and $z_{2}$. This identification relies on the
free field computation of the four point functions. One finds in particular
that for the contour of integration going between the two intermediate vertices
the block obtained is ${\cal F}^{1}-{\cal F}^{2}$, with no logarithm. It must
therefore correspond to the topology of figure 10. In figure 9, which
represents the "generic" situation, braiding of $z_{1}$ and $z_{2}$ results in
a non trivial operation, which we correlate with the presence of logarithms.

We now see that the scalar product $<en/I|e'n'/I>$ corresponds  to figure 11,
$<en/I|e'n'/II>$ to figure 12. In both case we still have to let $e_{0}
\rightarrow 0$
to get the partition function of $S^{2}\times S^{1}$ with two strands carrying
two dimensional representations. On the other hand if we consider
$<en/II|e'n'/II>$, this is represented as figure 13. Such invariant can be
computed eg by surgery, and one gets in $S^{3}$ two linked strands plus a
dotted loop. This is a split link whose Alexander invariant vanishes.

We also can understand the value of $S_{n,n'}$. With regularization,
 this is in fact the invariant for the link of
figure 14 where the two circles carry $(n)$ and $(n')$ representations and are
connected by a pair of lines carrying $(e_{0}n_{0})$ and $(-e_{0},1-n_{0})$.
 This
is the same as having a single loop carrying $(e_{0}n_{0})$, with partition
function ${\cal Z}=V/2\mbox{sin}(\pi e_{0}/k)$ that diverges as
$e_{0}\rightarrow 0$.

To check further the consistency, especially the role of the anomalous $T$
transformation, let us discuss the example of two unlinked circles in $S^{3}$,
one carrying $(en/I)$ and the other $(e'n'/I)$. Suppose first $e+e'\neq 0$.
Consider a solid torus $D^{2}\times S^{1}$ containing these two lines. The
state on its surface is
\[
|e+e',n+n'-1/I>-|e+e',n+n'/I>
\]
To compute the invariant we introduce a third line carrying a one dimensional
representation as in figure 15 and make a modular transformation.
 Since the modular matrix for $en/I\rightarrow\widehat{n''}$ does not
depend on $n$, the contributions of the two states cancel and we get a
vanishing result.

Suppose now $e+e'=0$. Then the state on the surface of the torus containing the
two loops is $\widehat{n+n'}$. By modular transformation it branches only to
states orthogonal to $(n'')$ representations, so we again get zero.

Now
consider a Hopf link with one strand carrying $(en)$, the other $(-e,n')$.
 Interpret this as the result of
$\tau\rightarrow\tau+1$ on the preceding case. Under this operation
\[
|\widehat{n+n'}>\rightarrow\mbox{exp}(-i\pi/6)\left(|\widehat{n+n'}>-
2i\epsilon V\sum_{n''}|n''>\right)
\]
Now apply $S$, and compute scalar product with the second torus containing a
one dimensional representation. Only the infinitesimal amount of $|n''>$ above
contributes, however the corresponding $S$ matrix element is infinitely large
so we get a scalar product simply equal to $iV\mbox{exp}(-i\pi/6)$.
We still have to correct for the change of framing in each of the circles when
$\tau\rightarrow\tau+1$, which results in a phase factor $\mbox{exp}[-2i\pi
(h_{en}+h_{-en'})]$, and the change of framing of $S^{3}$. Hence
\[
{\cal Z}=iV\mbox{exp}-2i\pi[e^{2}/k^{2}+(n-n')e/k]=S_{en/I,-en'/I}
\]
as it should be.

\subsection{Fusion rules}

The fusion rules \footnote{We use the convention $\Phi_{a}\cdot\Phi_{b}=\sum
N_{ab}^{c}\Phi_{c}$} can be immediately deduced from the above analysis and
\cite{RS91}. In fusion, two dotted lines extremities
associated with one of each representation must be attached, so two dotted
lines extremities remain free (see figure 16). We see in particular that a
 representation
$(\widehat{n})$ has regulators with similar position as $(en/I)$.
\[
(en/I)\cdot(e'n'/I)=(e+e',n+n'-1/I)-(e+e',n+n'/I)
\]
\[
(en/II)\cdot(e'n'/II)=(e+e',n+n'-1/II)-(e+e',n+n'/II)
\]
\[
(en/I)\cdot(e'n'/II)=(e+e',n+n'-1/I)-(e+e',n+n'/I)
\]
\[
(en/I)\cdot(-e,n'/I)=(\widehat{n+n'-1})\footnote{Notice that there is no
contribution from $(n+n'-1)$, although this representation appears in the
operator product $\Phi_{en}\Phi_{-e,n'}$. This is an example of the possible
differences between operator products and fusion rules in the Verlinde sense,
which is manifest also in some of the $N$ coefficients being greater than one.
An argument in favor of this fusion rule is that the log no log coupling in the
four point function is a purely quantum effect that disappears in the large $k$
limit. It is consistent with all further computations.}
\]
\[
(en/II)\cdot(-e,n'/II)=-2(n+n'-1)+(n+n')+(n+n'-2)
\]
\[
(en/I)\cdot(-e,n'/II)=(\widehat{n+n'-1})
\]
\[
(en/I)\cdot(\widehat{n'})=-2(e,n+n'/I)+(e,n+n'+1/I)+(e,n+n'-1/I)
\]
\[
(en/II)\cdot(\widehat{n'})=-2(e,n+n'/I)+(e,n+n'+1/I)+(e,n+n'-1/I)
\]
\[
(\widehat{n})\cdot(\widehat{n'})=-2(\widehat{n+n'})+
(\widehat{n+n'+1})+(\widehat{n+n'-1})
\]
\begin{equation}
(*n)\cdot(n')=(*n+n')
\end{equation}
where $(*n)$ means any representation with a $N$ number equal to $n$.

\subsection{An alternative basis}

To describe the fusion rules it sometimes is more convenient to introduce the
linear
combinations
\begin{equation}
|en/1>=|en/I>,\ |en/2>=|en/I>-|en/II>
\end{equation}
The second  partition function can be obtained as in section 3, by integrating
the screening operator along the contour $C_{6}$,
and one has
\begin{equation}
<en/1|e'n'/2>=0,\ <en/2|e'n'/2>=-\delta_{e+e'}\delta_{n+n'-1}
\end{equation}
Fusion of $(en/2)$ and $(-e,n'/2)$ produces a representation
$(\widehat{n+n'}/2)$, setting $(\widehat{n})=(\widehat{n}/1)$, and one has
\begin{equation}
\zeta_{\widehat{n},2}=\zeta_{\widehat{n}}+2\zeta_{n}-\zeta_{n+1}
-\zeta_{n-1}
\end{equation}
together with
\begin{equation}
<\widehat{n}/1|\widehat{n'}/2>=0,\
<\widehat{n}/2|\widehat{n'}/2>=2\delta_{n+n'}
-\delta_{n+n'-1}-\delta_{n+n'+1}
\end{equation}

Fusion rules do not mix species of type $1$ and type $2$. They look identical
for these two species up to signs.

\subsection{Factorization}

We must discuss briefly the factorization formulas that play a key role in the
computation of links and 3-manifolds invariants \cite{W89}. The case of
$gl(1,1)$ presents some difficulties related to the fact, explained above, that
for a sphere $S^{2}$ with a representation $(n)$ and its conjugate $(1-n)$
inserted the space of conformal blocks has vanishing dimension. Hence we cannot
a priori establish a connection between the invariant of say the sphere $S^{3}$
with two split links $L_{1}$, $L_{2}$ and the product of invariants of $S^{3}$
with either $L_{1}$ or $L_{2}$. To solve this difficulty, as in \cite{RS91}, we
can introduce a regularization by considering the link of figure 17 where the
dotted lines carry again $e_{0},n_{0}$. Assuming
\begin{equation}
\mbox{lim}_{e_{0}\rightarrow 0}{\cal Z}\left(\begin{array}{ccc}
\ &\ &\ \\
\ &\ &\ \\
\ &\ &\
\end{array}\right)={\cal Z}\left(\begin{array}{cc}
\ &\ \\
\ &\
\end{array}\right)
\end{equation}
we deduce from the same arguments as in \cite{W89}
\begin{equation}
{\cal Z}(S^{3};L_{1},L_{2})=\frac{{\cal Z}(S^{3};L_{1})\times
{\cal Z}(S^{3};L_{2})}{{\cal Z}(S^{3})}
\end{equation}
where ${\cal Z}(S^{3})$ is obtained from the invariant of $S^{3}$ with a loop
 carrying
$(e_{0}n_{0})$, as $e_{0}\rightarrow 0$. Using the above results we know that
this quantity is infinite, and its  regularized value is $V/2\epsilon$, as
$S_{nn'}$. Therefore set
\begin{equation}
{\cal Z}(S^{3})=\infty (reg.:V/2\epsilon)
\end{equation}
Now if both $L_{1}$ and $L_{2}$ had finite invariants, the invariant of
their disconnected sum vanishes, a result well known in the theory of the
Alexander Conway invariant. Suppose now that $L_{1}$ is say a loop carrying a
representation $(n)$ (a shadow component \cite{RS91}). On the one hand we
 expect then
\[
{\cal Z}(S^{3};L_{1},L_{2})={\cal Z}(S^{3};L_{2})
\]
on the other hand we get from the factorization formula
\[
{\cal Z}(S^{3};L_{1},L_{2})=\frac{{\cal Z}(S^{3};L_{2})\times{\infty}}{\infty}
\]
This indefinite ratio is well determined in the regularized version since then
${\cal Z}(S^{3};L_{1})=\infty(reg.:V/2\epsilon)={\cal Z}(S^{3})$, and it is
 simply equal to ${\cal Z}(S^{3};L_{2})
$, as desired.

\subsection{Verlinde formula}

We can now consider the Verlinde formula \cite{V88,MS90}, which from the knot
 theory point of
view, is the consistency equation for computing invariants of links as in
figure
18. Provided there is only one conformal block for the sphere $S^{2}$ with
insertion of a representation $\rho_{i}$ and its conjugate one has
\begin{equation}
S_{\rho_{i}\rho_{j}}S_{\rho_{i}\rho_{k}}/S_{0\rho_{i}}=\sum_{m}S_{\rho_{i}
\rho_{m}}N_{\rho_{j}\rho_{k}}^{\rho_{m}}
\end{equation}
Let us consider this formula for various cases of representation $\rho_{i}$.

First suppose $\rho_{i}=(n_{i})$, $\rho_{j}$ and $\rho_{k}$ are two dimensional
representations. Then the left member vanishes, corresponding to the vanishing
of a split link as discussed above. If $\rho_{j}$ and $\rho_{k}$ are of type
$II$, the right member vanishes term by term. If $\rho_{j}$ and $\rho_{k}$ are
of type $I$, two terms contribute. Suppose $\rho_{j}=(e_{j}n_{j}/I)$ and
$\rho_{k}=(e_{k}n_{k}/I)$. Then the terms
$\rho_{m}=(e_{j}+e_{k},n_{j}+n_{k}/I)$
and  $\rho_{m}=(e_{j}+e_{k},n_{j}+n_{k}-1/I)$ contribute, with opposite $N$
coefficients. However $S_{e'n'/I,n}$ does not depend on $n'$. Therefore the two
terms add to zero. Still for $\rho_{i}=(n_{i})$, let us mention the case where
$\rho_{j}=(n_{j})$, and say $\rho_{k}=(e_{k}n_{k})$.
 In that case the two infinities in the left hand side,
 regularized by the above $\epsilon$, cancel and one gets simply
$S_{e_{k}n_{k}/I,n_{i}}$. On the right hand side only one term contributes,
$\rho_{m}=(e_{k},n_{j}+n_{k}/I)$. But by formulas (\ref{eq:Hopf}),
\[
S_{e_{k}n_{k}/I,n_{i}}=S_{e_{k},n_{j}+n_{k}/I,n_{i}}
\]
so both things coincide again.

The most interesting situation arises when all of them are two dimensional
representations: $e_{i}n_{i}/I$, $e_{j}n_{j}/I$, $e_{k}n_{k}/I$. The left
hand side reads
\[
2V\mbox{sin}(\pi e_{i}/k)
\mbox{exp}\frac{2i\pi}{k}\left[e_{i}(\widetilde{n_{j}}+\widetilde{n_{k}}-1)+
(e_{j}+e_{k})(\widetilde{n_{i}}-1/2)\right]
\]
Let us now consider the right hand side. Because of graded fusion coefficients
it reads
\[
iV\mbox{exp}\frac{2i\pi}{k}\left[e_{i}(\widetilde{n_{j}}+\widetilde{n_{k}}-3/2)
+(e_{j}+e_{k})(\widetilde{n_{i}}-1/2)\right]
\]
\[
-iV\mbox{exp}\frac{2i\pi}{k}\left[e_{i}(\widetilde{n_{j}}+
\widetilde{n_{k}}-1/2)
+(e_{j}+e_{k})(\widetilde{n_{i}}-1/2)\right]
\]
which is
\[
iV[\mbox{exp}(-i\pi e_{i}/k)-\mbox{exp}(i\pi e_{i}/k)]
\mbox{exp}-\frac{2i\pi}{k}\left[e_{i}(\widetilde{n_{j}}+\widetilde{n_{k}}-1)
+(e_{j}+e_{k})(\widetilde{n_{i}}-1/2)\right]
\]
equal to the left hand side.

Therefore the presence of sine functions, which is a characteristic of the
multivariable Alexander Conway polynomial, occurs in some part of the formalism
due to infinite sums of exponential functions, in other parts because of
graded fusion rules.

Let us emphasize that the Verlinde formula is not expected to hold if
$\rho_{i}=(\widehat{n})$. In that case indeed the space of conformal blocks of
$S^{2}$ with $(\widehat{n})$ and its conjugate inserted has dimension greater
than one. Factorization arguments to cut the loop that carries $(\widehat{n})$
can therefore not be used. For the link represented in figure 19 for instance,
 one finds a vanishing
invariant. Similarly Verlinde formula is not expected to hold if
$\rho_{i}=(en/II)$ since then this space of conformal blocks has vanishing
dimension.

Let us also discuss briefly the Verlinde operators. The one corresponding to
the $b$ cycle, which we denote $V_{\rho_{i}}^{1,0}$ has its matrix
 elements readily
evaluated
\begin{equation}
\left[V_{\rho_{i}}^{1,0}\right]_{\rho_{j}}^{\rho_{k}}=
N_{\rho_{i}\rho_{j}}^{\rho_{k}}
\end{equation}
where the fusion coefficients $N$ have been given above. The one corresponding
to the $a$ cycle $V_{\rho_{i}}^{0,-1}$ is then obtained by modular
 transformation
\begin{equation}
V_{\rho_{i}}^{0,-1}=S^{-1} V_{\rho_{i}}^{1,0} S
\end{equation}
The operator
corresponding to a torus knot of type $(p,q)$ \cite{BZ85}
can be obtained via the formulas
\begin{equation}
V_{\rho_{i}}^{p,q}=S^{-1} V_{\rho_{i}}^{-q,p} S,\ V_{\rho_{i}}^{p,q}=
T^{-1} V_{\rho_{i}}^{p,q+p} T\label{eq:torusops}
\end{equation}

\subsection{Cabling in Alexander theory}

The composition rules of $gl(1,1)$ representations allow simple derivation of
the cabling formulas for Alexander invariants. Consider first some knot $K$
carrying some representation $(e_{1}n_{1})$. Double it by running parallel to
the original strand  another one carrying $(e_{2}n_{2})$ (figure 20). The
partition function of this system writes, due to (\ref{eq:phase})
\[
{\cal Z}=\left[\mbox{exp}(2i\pi w h_{e_{1}+e_{2},n_{1}+n_{2}-1})-
\mbox{exp}(2i\pi w
h_{e_{1}+e_{2},n_{1}+n_{2}})\right]\Box'_{e_{1}+e_{2}}(K)
\]
where $\Box'_{e_{1}+e_{2}}(K)$ is the invariant of $K$ carrying a
representation
with $E$ number $e_{1}+e_{2}$, $w$ is the writhe, sum
of the signs of the crossings of $K$. On the other hand, also by
 (\ref{eq:phase}),
\[
{\cal Z}=\mbox{exp}2i\pi
w\left[(n_{1}+n_{2}-1)(e_{1}+e_{2})/k+(e_{1}+e_{2})^{2}/2k^{2}\right]\Box'_{
e_{1},e_{2}}(\mbox{double of }K)
\]
from which one deduces immediately
\begin{equation}
\Box'_{e_{1},e_{2}}(\mbox{double of }K)=-2i
 \mbox{ sin}\pi w(e_{1}+e_{2})/k
\ \Box'_{e_{1}+e_{2}}(K)
\end{equation}
It is interesting to consider the limit $e_{1}+e_{2}\rightarrow 0$. In that
case it is well known (this can be deduced from the skein relation for the
ordinary Alexander Conway polynomial) \cite{K83} that
\[
\Box'_{e_{1}+e_{2}}(K)\approx -\frac{V}{2\mbox{sin}\pi(e_{1}+e_{2})/k}
\]
and therefore one finds that for a double knot with same $E$ numbers  but
opposite orientations (figure 21)
\begin{equation}
\Box'=iV w,\ \Delta=w
\end{equation}
One can also allow the second strand to wind around the first one.
 If the total twist
(figure 22) is t one finds, owing to the well known formula \cite{K83,BZ85}
\begin{equation}
l=w+t
\end{equation}
where $l$ is the linking number, the general result for a twisted double
\begin{equation}
\Box'_{e_{1}e_{2}}(\mbox{twisted double of }K)=-2i
 \mbox{ sin}\pi l(e_{1}+e_{2})/k\ \Box'_{e_{1}+e_{2}}(K)
\end{equation}
These formulas generalize easily to an $n$ cable (figure 23)
\begin{equation}
\Box'_{e_{1},e_{2},\ldots,e_{n}}(n\mbox{ cable of }K)=
\left(-2i\mbox{ sin}\pi w\sum
e_{i}/k\right)^{n-1}\Box'_{\sum e_{i}}(K)
\end{equation}

\subsection{Invariants of torus links}

Consider first a torus knot of type $(p,q)$ ($p,q$ being coprimes). We first
 claim that

\[
V_{en/I}^{p,q}|e'n'/I>=
\]
\begin{equation}
-\mbox{exp}\left[2i\pi\frac{q}{p}
(h_{e'+pe,n'+pn}-h_{e',n'})\right]|e'+pe,n'+pn/I>+\left[(en)\rightarrow
(e,n-1)\right]\label{eq:torusstate}
\end{equation}
By applying $T$ (which is trivial)
and $S$, one checks that each of the two terms in this sum has the right
behaviour (\ref{eq:torusops}). The precise combination can be found by
 considering the case
$p=1,q=0$. Now to obtain action of $V^{p,q}$ on a one dimensional
representation
we use formula (\ref{eq:prime}). Due to the minus sign in the above formula,
only
$p$ terms remain and we get
\begin{equation}
V_{en}^{p,q}|0>=\sum_{j=0}^{p-1}\mbox{exp}\left[2i\pi\frac{q}{p}
h_{pe,pn-j}\right]
|pe,pn-j/I>
\end{equation}
Using the value of $S_{0}^{en/II}$ we get the partition function
\begin{equation}
{\cal Z}(S^{3},(p,q))=\frac{-V}{2\mbox{sin }\pi pe/k} \frac{\mbox{sin}\pi
pqe/k}{\mbox{sin}\pi qe/k}\mbox{exp}(2i\pi pq h_{en})
\end{equation}
and therefore we get the general result for a torus knot, after appropriate
framing correction,
\begin{equation}
\Delta= -\frac{\mbox{sin}\pi pqe/k}{2i\mbox{ sin}(\pi pe/k)\mbox{sin}(\pi
qe/k)}
\end{equation}
in agreement with known formulas \cite{BZ85}.

We can with the same formalism deal with torus links of type $(ap,aq)$, $q>1$.
In that case (\ref{eq:torusstate}) becomes
\[
V^{ap,aq}_{en/I}|e'n'/I>=
\]
\begin{equation}
(-)^{a}\sum_{j=0}^{a}(-1)^{j}\left(\begin{array}{c}
j\\
a
\end{array}\right)\mbox{exp}\left[2i\pi\frac{q}{p}(h_{e'+ape,n'+apn-pj}
-h_{e'n'})\right]|e'+ape,n'+apn-pj/I>
\end{equation}
the combination of representations being found by considering the case
 $p=1,q=0$. We get therefore
\begin{equation}
\Delta=(-2i)^{a-2}\frac{(\mbox{sin}\pi apqe/k)^{a}}
{\mbox{sin}(\pi ape/k)\mbox{sin}(\pi
aqe/k)}
\end{equation}

\section{3-Manifold Invariants}

\subsection{$X_{h}\times S^{1}$}

We have already encountered the following
\begin{equation}
{\cal Z}(S^{3})=\infty,\ {\cal Z}(S^{2}\times S^{1})=0
\end{equation}
the last result occuring because there is no conformal block on the sphere with
one dimensional representations inserted, or $<n|n'>=0$. Consider now the torus
$S^{1}\times S^{1}$. Its Hilbert space contains twice as many fields as the
volume of the fundamental domain due to the doubling of representations
explained above. We therefore expect
\begin{equation}
{\cal Z}(S^{1}\times S^{1}\times S^{1})=2V^{-2}
\end{equation}
Consider now a 3-manifold of the form $M=X_{h}\times S^{1}$ where $X_{h}$
 is a
surface of genus $h$. It can be obtained by surgery on the system of loops of
figure 24. For an ordinary theory like $SU(n)$, as discussed in \cite{WW89},
one finds the result
\begin{equation}
\sum_{i}(S_{0\rho_{i}})^{1-h}(S_{0}^{\rho_{i}})^{1-h}
\end{equation}
Suppose we forget for a while subtleties attached with vanishing values of the
$e$ number and compute this sum by taking the expressions of the $S$ matrix
elements established in the case of two dimensional representations and $e\neq
0$. Then one finds for $(en/1)$
\begin{equation}
V^{2-2h}\sum_{en} [\mbox{exp}(i\pi e/k)-\mbox{exp}(-i\pi e/k)]^{2h-2}=V^{-2h}
\left(\begin{array}{c}
h-1\\
2h-2
\end{array}\right)
\end{equation}
and, up to a sign, an identical result for $(en/2)$ representations (this
combinatorial factor can also be obtained by counting paths on the $gl(1,1)$
Bratteli diagram \cite{RS91}). Therefore
 we
expect
\begin{equation}
{\cal Z}(X_{h}\times
S^{1})=\frac{(2h-2)!}{[(h-1)!]^{2}}\left[1+(-)^{h-1}\right]
V^{-2h}
\end{equation}
It is better to explicitely check this result by manipulating our various
 representations
and taking correctly into account modular properties.
We will first consider $S^{1}\times S^{1}\times
S^{1}$. This three dimensional torus can be produced from $S^{3}$ by surgery on
the link of figure 25. Let us first do surgery on loop 1, to get $S^{2}\times
S^{1}$ with loops 2 and 3 linked in it as in figure 26. Now instead of
performing an actual surgery on loop 2 we can consider it as a Wilson loop
carrying all possible representations $\rho$ along it, each one coming with a
factor $S_{0}^{\rho}$. We then have to compute a trace of Chern Simons
evolution operator between the states in sections 1 and 2.
Suppose first $\rho$ is of type $(en/II)$. There is then only one conformal
block for sections 1 and 2. One can therefore using factorization arguments
perform the calculation by glueing caps as in figure 27 and dividing by the
 scalar
product of caps themshelves, which is clearly $S^{3}$ with the unknot carrying
$(en/II)$. We finally perform surgery on loop 3 to obtain an $S^{2}\times
S^{1}$ that contains two Wilson loops carrying $(en/II)$ with opposite
orientations. Therefore we find the contribution of a given $(en/II)$
representation
\[
\frac{S_{0}^{en/II}}{S_{0,en/II}}<en/II|C|en/II>
\]
Unfortunately this is an undeterminate since the denominator vanishes as well
as
the scalar product. To get a finite result we must change the basis. A natural
basis to consider is the representations $|en/1>,|en/2>$. In that case one has
$S_{0}^{en/1}=-S_{0}^{en/2} =S_{0,en/1}=S_{0,en/2}$ and thus one finds a
factor of 2 per representation $(en/II)$. Therefore the contribution of $\rho$
 irreducible to ${\cal Z}(S^{1}\times S^{1}\times S^{1})$ is
\[
2\times 2k(2k-1)
\]
Let us now discuss the contribution originating from $\rho=(\widehat{n})$
an indecomposable block. The
situation is more complicated because now there are two conformal blocks for
sections 1 and 2. One block is logarithmless, let us call it $|1>$, and the
other contains logarithms, let us call it $|2>$ (it is well defined only up to
addition of $|1>$). The partition function of $S^{2}\times S^{1}$ as
in figure 26 can be expressed as the trace of some matrix ${\cal M}$.
Let us replace the loop carrying
an indecomposable block representation by two parallel loops of opposite
orientation, one carrying
$(e,n_{1}/I)$ and the other $(e,n_{2}/I)$. Then the "body" looks as figure 28
 while there are two
possible caps extracting the conformal block $|1>$ (figure 29) or $|2>$ (figure
30). Now $<1|1>=<\mbox{cap}1|\mbox{cap}1>$ and $<2|2>$ are the Alexander
 polynomials of
$S^{3}$ with two
disconnected loops, which are known to vanish. Similarly
$<1|2>=<\mbox{cap}1|\mbox{cap}2>$ is the Alexander polynomial of $S^{3}$ with
the unknot, with value $S_{0,en/I}$. Let us consider now the quantity
\begin{equation}
\frac{<1|{\cal M}|2>}{<1|2>}={\cal M}_{22}=
\frac{<2|{\cal M}|1>}{<2|1>}={\cal M}_{11}
\end{equation}
The numerator of this equation is the partition function of the body of figure
28 with cap 1 glued on top and cap 2 on the bottom. The result is presented in
figure 31 (dotted lines originating from regularization). By factorization
 we get
\begin{equation}
{\cal Z}(\mbox{figure} 31)=\frac{{\cal Z}(\mbox{figure} 32)
S_{0,en/I}}{{\cal Z}(S^{3})}
\end{equation}
The factor ${\cal
Z}(S^{3})$ that diverges when $\epsilon\rightarrow 0$ will just compensate
 for the $S_{0}^{\widehat{n}}$
generated when putting $(\widehat{n})$ on loop 2. We are thus left with
computing ${\cal Z}(\mbox{figure} 32)$. By surgery on loop 3 we get
$S^{2}\times S^{1}$ with only one conformal block, so
\[
{\cal M}_{22}=1
\]
Therefore
\begin{equation}
{\cal M}_{11}={\cal M}_{22}=1,\ Tr{\cal M}=2
\end{equation}
so each block representation contributes a factor 2. We thus confirm
by this explicit computation the doubling phenomena.

As for $X_{h}\times S^{1}$, which can be obtained by surgery on the loops of
figure 24, a similar computation can be done.

\subsection{Seifert Manifolds}

We now would like to discuss Seifert manifolds
\[
X(p_{1}/q_{1},\ldots,p_{n}/q_{n})
\]
where $p_{i}$ and $q_{i}$ are coprimes. Recall that $X$ is obtained from
$S^{2}\times S^{1}$ by removing $n$ disjoint solid tori $D_{i}\times S^{1}$
($D_{1},\ldots,D_{n}$ disjoint disks in $S^{2}$) and glueing them back after
twisting the boundary by certain $SL(2,Z)$ matrices $M_{1},\ldots,M_{n}$. The
matrix $M_{i}=M(p_{i},q_{i})$ has first column $\left(\begin{array}{c}
p_{i}\\
q_{i}
\end{array}\right)$. The particular choice of $M_{i}$ affects the 2-framing,
but not the diffeomorphism type of the resulting manifold (see \cite{FG90}).

Let us first consider the Lens space $L(q,p)=X(p/q)$. First of all notice that
in a basis $|n>,|\widehat{n'}>$ we can write
\begin{equation}
S=\left(\begin{array}{cc}
0&-2\epsilon \\
-1/2\epsilon&0
\end{array}\right)
\end{equation}
and
\begin{equation}
T=\mbox{exp}(-i\pi/6)\left(\begin{array}{cc}
1&-2i\epsilon\\
0&1
\end{array}\right)
\end{equation}
therefore
\begin{equation}
T^{p}S=i\ \mbox{exp}(-i\pi/6)\left(\begin{array}{cc}
p&2i\epsilon\\
i/2\epsilon&0
\end{array}\right)
\end{equation}
in the sense that
\begin{equation}
T^{p}S\left(a|n>+b|\widehat{n'}>\right)=i\ \mbox{exp}(-i\pi/6)V\sum_{mm'}
\frac{i}{2\epsilon}a |\widehat{m}>+(2i\epsilon b+p) |m'>
\end{equation}
Let us now write a continued fraction expansion
\begin{equation}
p/q=a_{r}-\frac{1}{a_{r-1}-\frac{1}{\ldots-\frac{1}{a_{1}}}}
\end{equation}
then chose for $M_{i}$
\begin{equation}
M_{i}=\left(\begin{array}{cc}
a_{r}&-1\\
1&0
\end{array}\right)\left(\begin{array}{cc}
a_{r-1}&-1\\
1&0
\end{array}\right)\left(\begin{array}{cc}
a_{1}&-1\\
1&0
\end{array}\right)
\end{equation}
with
\begin{equation}
\left(\begin{array}{cc}
a&-1\\
1&0
\end{array}\right)=T^{a}S
\end{equation}
Now let us suppose for the moment that $T,S$ transformations restricted to the
space of $|n>,|\widehat{n'}>$ give the invariant of the Lens space. In that
case, starting with a representation $|0>$ and making appropriate surgery gives
\begin{equation}
i^{r}\mbox{exp}(-i\frac{\pi}{6}\sum_{j=1}^{j=r} a_{j})
\prod_{j=r}^{j=1}\left(\begin{array}{cc}
a_{j}&2i\epsilon\\
i/2\epsilon&0
\end{array}\right)\left(\begin{array}{c}
V\\
0
\end{array}\right)
\end{equation}
from which one deduces
\begin{equation}
{\cal Z}(L(q,p))=i^{r}\mbox{exp}(-i\frac{\pi}{6}\sum_{j=1}^{r}a_{j})
\frac{Viq}{2\epsilon}
\end{equation}
It is easy to check that apart from  $|n>,|\widehat{n'}>$, the other
representations indeed do not contribute to the final result due to summation
 over
all possible $n$ numbers. Recall that $L(0,1)=S^{2}\times S^{1},
 L(1,0)=S^{3}$,
in agreement with already established results.

For a  Seifert manifold $X(p_{1}/q_{1},\ldots,p_{n}/q_{n})$ one finds
immediately
\begin{equation}
{\cal Z}(X(p_{1}/q_{1},\ldots,p_{n}/q_{n}))=\prod_{i=1}^{n}
i^{r_{i}}\mbox{exp}(-i\frac{\pi}{6}\sum_{j=1}^{r_{i}}a_{ij})\frac{V}{2\epsilon}\
[ip_{1}p_{2}\ldots p_{n-1}q_{n}+\mbox{permutations}]\label{eq:seifert}
\end{equation}

Strictly speaking we still have to let $\epsilon\rightarrow 0$ so all these
invariants become infinite. However the ratio ${\cal Z}(X)/{\cal Z}(S^{3})$
remains well defined. The necessary corrections for framing are discussed in
details in \cite{FG90}. After they are taken into account, the invariant
obtained depends only trivially on the level $k$ by an overall scale. The non
trivial content is $k$ independent and reduces essentially to the absolute
value of
\begin{equation}
p_{1}p_{2}\ldots p_{n-1}q_{n}+\mbox{permutations}
\end{equation}
which is nothing but the order (number of elements) of the first homology group
   of $X$ (with the
convention that zero indicates infinite order).

Recall that  $X(-2/1,3/1)\approx S^{3}$ and
\begin{equation}
X(p_{1}/q_{1},\ldots,p_{n}/q_{n},0/1)=\mbox{disconnected sum of }
L(-p_{i},q_{i})
\end{equation}
both results in agreement with  (\ref{eq:seifert}).

\subsection{A three dimensional point of view}

We now wish to discuss the divergence of the $S^{3}$ invariant
from the 3 dimensional Chern Simons point of view. Recall that in the large $k$
limit, ${\cal Z}(S^{3})$ can be estimated using the saddle points of the Chern
Simons integral, given by flat connections. For $S^{3}$ there is only the
trivial connection, so one gets for an arbitrary Lie group $G$
\begin{equation}
{\cal Z}(S^{3})\propto \frac{1}{\mbox{Vol}(G)},\ k\rightarrow \infty
\end{equation}
(It is necessary to divide by the volume of the group as the connection is
reducible and is invariant under global gauge transformations). In the $SU(n)$
 case for instance one has
\begin{equation}
\mbox{Vol}(SU(n))=\prod_{p=2}^{p=n}\frac{2\pi^{p}}{\Gamma(p)}
\end{equation}
in agreement with
\begin{equation}
S_{00}\approx
(2\pi)^{n(n-1)/2}\frac{1}{k^{(n-1)(n+1)/2}}\frac{\Gamma(n-1)\ldots\Gamma(2)}
{\sqrt{n}}
\end{equation}
Notice that if one naively lets $n\rightarrow 0$ in that equation one finds
$S_{00}\rightarrow \infty$.

We now discuss the volume of $U(1,1)$. The most direct way to find the
appropriate measure is to study characters. We write an element $g$ of $U(1,1)$
as
\begin{equation}
g=\mbox{exp}i(x_{e}N+x_{n}E)\mbox{exp}i(x_{\psi^{+}}\psi+x_{\psi}\psi^{+})
\end{equation}
Consider first a representation
$(en)$. Then one has
\begin{equation}
\mbox{ch}_{en}(g)=Str_{en}(g)=\mbox{exp}i[x_{n}e+x_{e}(n-1/2)]\left[
e^{-ix_{e}/2}(1-\frac{e}{2}x_{\psi^{+}}x_{\psi})-e^{ix_{e}/e}(1+
\frac{e}{2}x_{\psi^{+}}
x_{\psi})\right]
\end{equation}
We discard now the $x_{\psi}x_{\psi^{+}}$ contribution for simplicity, ie
restrict to $g$ belonging to the maximal torus. Then
\begin{equation}
\mbox{ch}_{en}(g)=-2i\mbox{ exp}i[x_{n}e+x_{e}(n-1/2)]\mbox{sin}(x_{e}/2)
\end{equation}
and also
\begin{equation}
\mbox{ch}_{\widehat{n}}(g)=-4e^{ix_{e}n}\mbox{sin}^{2}(x_{e}/2)
\end{equation}
\begin{equation}
\mbox{ch}_{n}(g)=e^{ix_{e}n}
\end{equation}
The range of parameters is $x_{e}\in [0,4\pi]$, $x_{n}\in [0,2\pi]$. It is a
very natural assumption that the representations $(en)$ and $(-e,1-n)$ are
conjugate. Calling $d\mu$ the measure of integration for class functions we
require therefore
\begin{equation}
<en|e'n'>=\int d\mu
\mbox{ ch}_{en}(g)\mbox{ch}_{e'n'}(g)=\delta_{e+e'}\delta_{n+n'-1}
\end{equation}
so
\begin{equation}
d\mu=-\frac{dx_{e}dx_{n}}{4\mbox{sin}^{2}(x_{e}/2)}
\end{equation}
This form is the naive extension of well known results in the theory of
ordinary Lie groups where the measure for class functions contains
$\mbox{sin}^{2}$ from each of the roots. In the supergroup case fermionic roots
, due to the standard formulas for change of variables in Grassman integration
\cite{deWitt87},
provide the inverse of such terms \cite{Kac}. Now for $<n|n'>$ we find
\begin{equation}
<n|n'>=-\int dx_{e}dx_{n}\frac{e^{ix_{e}(n+n')}}{4\mbox{sin}^{2}(x_{e}/2)}
\end{equation}
To give a meaning to this integral we give to $x_{e}$ a small imaginary part so
that
\begin{equation}
<n|n'>=2\pi\int_{0}^{4\pi} dx_{e}e^{ix_{e}(n+n')}\sum_{j=0}^{\infty}
e^{-i(2j+1)x_{e}}e^{-2j\epsilon}
\end{equation}
and each term gives vanishing contribution once integrated due to periodicity.
Therefore we recover $<n|n'>=0$ as claimed earlier in the text. Other scalar
products are recovered in a similar fashion. In particular we see now that
\begin{equation}
\mbox{Vol}(U(1,1))\propto \int d\mu\propto <0|0>=0
\end{equation}
hence reaching the remarkable result that $U(1,1)$ has a vanishing volume.

Once plotted in the large $k$ expansion, this vanishing volume explains the
divergence of the $S^{3}$ invariant. More generally such divergence should
occur for any 3-manifold that admit a discrete set of flat connections. The
divergence comes from the flat connections that commute with the
entire $gl(1,1)$, ie are of the form $A^{\mu}=A^{\mu}_{N}E$. In that case the
covariant derivative with respect to $A$ is the same as the ordinary
derivative, and the study of fluctuations around the saddle point involves only
Gaussian integrals. The corresponding determinants in numerator and denominator
cancel up to signs to due boson fermion symmetry. The result is then identical
to the invariant that would be obtained from super $IU(1)$, and as argued in
 \cite{Witten89} it should be nothing but the $U(1)$ Casson invariant. Indeed
the $U(1)$ Casson invariant counts the maps from $\Pi_{1}({\cal M})$ into
$U(1)$ (up to conjugacy). Since $U(1)$ is commutative this counts in fact maps
from $\frac{\Pi_{1}}{[\Pi_{1},\Pi_{1}]}({\cal M})$, ie $H_{1}({\cal M})$, into
$U(1)$. But this number of maps has to coincide with the number of elements of
$H_{1}$
since the latter is commutative.

\subsection{Invariants of links in 3-manifolds}

We have in principle at our disposal all the necessary ingredients to compute
invariants of links in 3-manifolds. As an example we consider again the lens
space $L(q,p)$. To put a knot in it we start with
$S^{2}\times S^{1}$, remove a  solid tori $D\times S^{1}$
that contains a Wilson loop carrying the representation $(en/I)$,
and glue it back after
twisting the boundary by a $SL(2,Z)$ matrix with
 first column $\left(\begin{array}{c}
p\\
q
\end{array}\right)$. We suppose moreover
\begin{equation}
p/q=a_{2}-\frac{1}{a_{1}}
\end{equation}
To compute the corresponding invariant we follow the general strategy, ie apply
$T^{a_{2}}ST^{a_{1}}S$ to $|en/I>$ and evaluate scalar product of the final
state with $|n=0>$. One finds the result
\[
{\cal
Z}(L(q,p),K)=
\]
\begin{equation}
-iV^{2}\mbox{exp}(-i\frac{\pi}{6}(a_{1}+a_{2}))\sum_{n'}\sum_{e'\neq
0}
\frac{\mbox{exp}\frac{2i\pi}{k}
[ae'(\widetilde{n'}-1/2)-e'(\widetilde{n}-1/2)-e(\widetilde{n'}-1/2)]}
{2\mbox{sin}\pi e'/k}
\end{equation}
The result depends therefore on the number of solutions in the fundamental
domain of the equation (for the unknown $e'$)
\begin{equation}
qe'=e \mbox{ mod}2k
\end{equation}
If there is no solution the invariant vanishes. Suppose there is a single
solution of the form $e'=e/q$. Then the invariant is
\begin{equation}
{\cal Z}(L(q,p),K)=-iV\mbox{exp}(-i\frac{\pi}{6}(a_{1}+a_{2}))\frac{\mbox{exp}
-\frac{2i\pi}{k}\frac{e}{q}(\tilde{n}-1/2)}{2\mbox{sin}\pi e/kq}
\end{equation}
In the limit $e=e_{0}\rightarrow 0$ one finds then
\begin{equation}
{\cal Z}(L(q,p),K)\rightarrow\
-\mbox{exp}(-i\frac{\pi}{6}(a_{1}+a_{2}))\frac{Viq}{2\epsilon}
\end{equation}
in agreement with the above computation of the  invariant of an "empty" Lens
space. In the case of $L(1,0)=S^{3}$ one also recovers the invariant of the
 unknot
is
$S^{3}$. In general the result depends on arithmetic properties of $e$ and $q$.

\section{Conclusion}

The study of the WZW on $GL(1,1)$ presented here, together with the results of
\cite{RS91}, show a complicated and rather unexpected pattern. Although we are
not totally happy with our derivations ($gl(1,1)^{(1)}$ characters
for instance need a more complete and rigorous analysis), be believe the
 results to be correct. In particular they reproduce nicely known all known
features of the multivariable Alexander Conway polynomial. Moreover we have
strong indication that the present pattern extends to other super groups, eg
$SU(n,m)$ with mixtures of typical, atypical and indecomposable blocks, and
that the picture presented in \cite{BMRS91} is too simple.

Although it may look a little surprising to find possibly infinite
3-manifold invariants, the emerging relation with homology (which is
expected from the classical theory of Alexander invariants \cite{BZ85}) makes
these infinities quite natural. Introduce ${\cal O}({\cal M})$ to be the
order of the first homology group of the 3-manifold ${\cal M}$. Take the
 convention
that this number is zero if the homology group is infinite. Then in all cases
we considered the following holds
\begin{equation}
\frac{{\cal Z}({\cal M})}{{\cal Z}(S^{3})}\propto{\cal O}({\cal
M})\label{eq:homo}
\end{equation}
(up to framing and normalization factors). Indeed for ${\cal M}=X_{h}\times
 S^{1}$, the order of the homology group is
infinite and this ratio is zero. For ${\cal M}$ a Seifert manifold the result
was obtained in the last section. Moreover, while ${\cal O}(S^{3})=1$, with
 order
one, if we put
inside a knot $K$ carrying a representation with non vanishing $E$ number, the
complement $S^{3}-K$ aquires  a first homology group equal to $Z$, with order
zero. This matches the fact that ${\cal Z}(S^{3})/{\cal Z}(S^{3})=1$ while
${\cal Z}(S^{3},K)/{\cal Z}(S^{3})=0$. Therefore our results are perfectly
consistent with topology. On the basis of the discussions in the last section
it is likely that (\ref{eq:homo}) is generally true and coincides with the
$U(1)$ Casson invariant, which can also be interpreted as a kind of
Reidemeister torsion \cite{Milnor}. We hope to get back to the topological
meaning of
Alexander invariants of links in 3-manifolds and their relation with
Reidemeister torsion.

\bigskip

{\bf Acknowledgments}: We thank M.Bershadsky, L.Crane, D.Freed, V.Kac,
 L.Kauffman, G.Moore, V.Serganova , V.Turaev and A.Vaintrob
 for useful discussions.

\pagebreak

{\bf Appendix 1}

The purpose of this appendix is to discuss briefly $U_{q}gl(1,1)$ in the
context of the general construction for 3-manifold invariants proposed in
\cite{RT91} for the non graded case. First of all introduce
\begin{equation}
K=q^{(E-N)/2},\ L=q^{(E+N)/2}
\end{equation}
one can then write the defining relation of $U_{q}gl(1,1)$ as
\begin{equation}
\left\{\psi,\psi^{+}\right\}=\frac{KL-(KL)^{-1}}{q-q^{-1}}
\end{equation}
and
\begin{equation}
K\psi^{(+)} K^{-1}=q^{(-)1/2}\psi^{(+)},\ L\psi^{(+)}
L^{-1}=q^{-(+)1/2}\psi^{(+)}
\end{equation}
The coproduct reads
\[
\Delta(K^{\pm})=K^{\pm}\otimes K^{\pm},\ \Delta(L^{\pm})=L^{\pm}\otimes L^{\pm}
\]
\begin{equation}
\Delta(\psi^{(+)})=\psi^{(+)}\otimes (KL)^{1/2}+(KL)^{-1/2}\otimes \psi^{(+)}
\end{equation}
the counit
\begin{equation}
\varepsilon(\psi)=\varepsilon(\psi^{+})=0,\ \varepsilon(K)=\varepsilon(L)=1
\end{equation}
and the antipode
\begin{equation}
s(K)=K^{-1},\ s(L)=L^{-1},\ s(\psi^{(+)})=-\psi^{(+)}
\end{equation}
The universal ${\cal R}$ matrix takes a very simple form
\begin{equation}
{\cal R}=\left[1+(q-q^{-1})q^{(E\otimes 1-1\otimes
E)/2}\psi^{+}\otimes\psi\right]q^{-E\otimes N-N\otimes E}=\sum_{i}\alpha_{i}
\otimes\beta_{i}
\end{equation}
where $\alpha$ and $\beta$ are obtained by expanding the exponentials.
Introduce now the element
\begin{equation}
u=\sum s(\beta_{i})\alpha_{i}=q^{2EN}\left[1+(q-q^{-1})q^{E}\psi\psi^{+}\right]
\end{equation}
with
\begin{equation}
s(u)=\sum\alpha_{i}
s(\beta_{i})=q^{2EN}\left[1-(q-q^{-1})q^{-E}\psi^{+}\psi\right]
\end{equation}
and
\begin{equation}
u^{-1}=\sum\beta_{i}s^{2}(\alpha_{i})
\end{equation}
One has
\begin{equation}
\epsilon(u)=\epsilon(s(u))=1
\end{equation}
Then for any  element $a$ in the algebra
\begin{equation}
uau^{-1}=s^{2}(a)
\end{equation}
Set
\begin{equation}
v^{2}=us(u)=s(u)u=
q^{4EN}\left[1+(q-q^{-1})(q^{E}\psi\psi^{+}-q^{-E}\psi^{+}\psi)\right]
\end{equation}
Then $v$ is central. Now consider a pair of representations $\rho\rho'$ and
 define
\begin{equation}
{\cal S}_{\rho\rho'}=\sum_{ij}
Str_{\rho}(uv^{-1}\beta_{i}\alpha_{j})
Str_{\rho'}(uv^{-1}\alpha_{i}\beta_{j})\label{eq:quantums}
\end{equation}
It is easy to check that when $\rho$ or $\rho'$ (or both) is a two or four
dimensional
representation, this vanishes. Such a result is well known in the theory of
the Alexander Conway polynomial where $S_{\rho\rho'}$ is (up to a framing
correction) the "bare" invariant for a Hopf link (figure 8), which vanishes
because the superdimension of $\rho$ and $\rho'$ is zero, or alternatively
because the Alexander matrix has a zero mode. If one sticks to
(\ref{eq:quantums}), a non vanishing result is obtained only for $\rho$ and
$\rho'$ one dimensional, but it is trivial.

To get interesting Alexander invariants, a strand has to be "open". This means
one has to take supertrace only on one of the representations, say $\rho'$. Let
us therefore define
\begin{equation}
{\cal S}_{\rho\rho'}=\sum_{ij}
Tr_{(b\mbox{ in }\rho)}(uv^{-1}\beta_{i}\alpha_{j})
Str_{\rho'}(uv^{-1}\alpha_{i}\beta_{j})\label{eq:quantums'}
\end{equation}
where the first factor is a trace restricted to bosonic states in $\rho$.
Suppose $\rho=(en)$, $\rho'=(e'n')$. Then
\begin{equation}
{\cal S}_{\rho\rho'}=q^{2e+e'}q^{-2(e'm+em')}\left(q^{e}-q^{-e}\right)
\end{equation}
Now, as discussed in \cite{RS91} to get link invariants when one opens a strand
it is necessary to rescale the above result by a factor depending on the $e$
number of the strand that has been open, leading to
\begin{equation}
S_{en,e'n'}=\frac{iV}{q^{2e}-1}{\cal
S}_{en,e'n'}=iV\mbox{exp}\frac{2i\pi}{k}[e'(n-1/2)+e(n'-1/2)]
\end{equation}
where we used $q=\mbox{exp}(-i\pi/k)$. Suppose now $\rho'$ is a one
 dimensional representation $(n')$. Then one finds
\begin{equation}
S_{en,n'}=-V\mbox{exp}(2i\pi en'/k)/2\mbox{sin}(\pi e/k)
\end{equation}
(If $\rho$ was one dimensional and $\rho'=(en)$ we would get an indeterminate
form). Suppose $\rho'$ is a four dimensional representation $(\widehat{n'})$.
Then
\begin{equation}
S_{en,\widehat{n'}}=2V\mbox{sin}(\pi e/k)\mbox{exp}(2i\pi en'/k)
\end{equation}
Up to the $1/k^{2}$ quantum corrections to the conformal weight of the fields,
these formulas agree with what we derived above from modular transformations.
$S_{nn'}$ is infinite  because of the normalization factor
, which on the other hand is necessary to provide link invariants.
Not opening strands gives trivial invariants, while opening them
 forces us to treat also one dimensional representations ("shadow components"
\cite{RS91}) and to face divergences.

However the quantum group approach does not seem to provide any insight as to
what are the type $II$ representations. This doubling still seems to be a
purely quantum effect.

\pagebreak

{\bf Appendix 2}

In this Appendix we discuss in more details the computation of the Alexander
polynomial for torus knots and links. Consider a strand which is wound
$p$ times along the noncontractible cycle and $q$ times along the
contractible cycle of a solid torus which belongs to $S^{3}$. If $p$
and $q$ are coprime and the torus is not knotted in $S^{3}$, then the
strand forms a torus knot of the type $(p,q)$. More generally we can consider
torus links with windings $(ap,aq)$ where $p$ and $q$ are coprimes. They
correspond to
 $a$ linked $(p,q)$ knots.

We regard a $(P,Q)$ torus knot (or link) as a Verlinde operator
 $V^{PQ}_{en}$. Here $|en>$ specifies a typical
$gl(1,1)$ representation flowing along the strand. We assume that
all typical representations appearing in this Appendix are of the type
$I$ and that the framing of the strand goes parallel (or
perpendicular) to the surface of the solid torus. We will derive the
action of $V_{en}^{PQ}$ on the Hilbert space of Kac-Moody
characters. This operator should satisfy the following identities when
conjugated with modular transformations $S$ and $T$:
\begin{eqnarray}
SV_{en}^{P,Q}S^{-1}&=&V_{en}^{-Q,P}
\label{A.1}\\
TV_{en}^{P,Q}T^{-1}&=&V_{en}^{P,Q+P}
\label{A.2}
\end{eqnarray}

It is easy to see that a repeated conjugation of $V_{en}^{P,Q}$ with
 $T$ and $S$ can reduce it to $V_{en}^{a,0}$ (where $a$
 is the greatest common divisor of $P$ and $Q$).
The action of Verlinde's operator $V_{en}^{a,0}$ can be
directly expressed through the fusion rules
\begin{equation}
V_{en}^{1,0}|e_{1},n_{1}>=
|e_{1}+e,n_{1}+n-1> -|e_{1}+e,n_{1}+n>
\label{A.3}
\end{equation}
and generally
\begin{equation}
V_{en}^{a,0}|e_{1},n_{1}>=(-)^{a}
\sum_{j=0}^{a}\left(
\begin{array}{cc}
j\\
a
\end{array}
\right)(-1)^{j}|e_{1}+ae,n_{1}+an-j>
\label{A.4}
\end{equation}

Our strategy is to propose  a formula for $V_{en}^{PQ}|e_{1},n_{1}>$
 and then to check that it satisfies
equations (\ref{A.1}) and (\ref{A.2}) as well (\ref{A.3}) or
(\ref{A.4}) in the case when $Q=0$. We will represent $V_{en}^{PQ}$
 as a linear combination of the elementary
operators $V_{*en}^{PQ}$, whose action is defined as
follows:
\begin{equation}
V_{*en}^{PQ}|e_{1},n_{1}>=
\mbox{exp}\left[2\pi i\frac{Q}{P}(h_{e_{1}+pe,n_{1}+pn}-h_{e_{1},n_{1}})\right]
|e_{1}+pe,n_{1}+pn>
\label{A.5}
\end{equation}
where the conformal weights
$h_{en}$ are given in the text.
The operator $V_{*en}^{PQ}$ satisfies both
eqs.~(\ref{A.1}) and (\ref{A.2}). Indeed,
\begin{eqnarray}
&&TV_{*en}^{PQ}T^{-1}|e_{1},n_{1}>=
\mbox{exp}\left[-\frac{2\pi i}{k}e_{1}(\widetilde{n_{1}}-1/2)\right]
TV_{*en}^{PQ}|e_{1},n_{1}>\nonumber\\
&&=-\mbox{exp}\frac{2\pi i}{kP}\left[Q(e_{1}+Pe)(\widetilde{n_{1}+Pn}-1/2)-
(Q+P)e_{1}(\widetilde{n_{1}}-1/2)\right]T|e_{1}+Pe,n_{1}+Pn>\nonumber\\
&&=-\mbox{exp}\frac{2\pi i}{kP}\left[(Q+P)(e_{1}+Pe)
(\widetilde{n_{1}+Pn}-1/2)-
(Q+P)e_{1}(\widetilde{n_{1}}-1/2)\right]|e_{1}+Pe,n_{1}+Pn>\nonumber\\&&
=V_{*en}^{P,Q+P}|e_{1},n_{1}>
\label{A.9}
\end{eqnarray}
and
\begin{eqnarray}
&&SV_{*en}^{PQ}S^{-1}|e_{1},n_{1}>=-iV\sum_{e_{2},n_{2}}
\mbox{exp}\frac{2\pi
i}{k}\left[e_{1}(\widetilde{n_{2}}-1/2)+e_{2}(\widetilde{n_{1}}-1/2)\right]
SV_{*en}^{PQ}|e_{2},n_{2}>\nonumber\\
&&=-iV\sum_{e_{2},n_{2}}\mbox{exp}\frac{2\pi
i}{kP}\left\{P[e_{1}(\widetilde{n_{2}}-1/2)+
e_{2}(\widetilde{n_{1}}-1/2)]
+Q(e_{2}+Pe)(\widetilde{n_{2}+Pn}-1/2)\right.\nonumber\\
&&\left.-Qe_{2}(\widetilde{n_{2}}-
1/2)\right\}S|e_{2}+Pe,n_{2}+Pn>\nonumber\\
&&=V^{2}\sum_{e_{3},n_{3}}\sum_{e_{2},n_{2}}\mbox{exp}\frac{2\pi
i}{kP}\left\{
-P[e_{3}(\widetilde{n_{2}+Pn}-1/2)-(e_{2}+Pe)(\widetilde{n_{3}}-1/2)]\right.
\nonumber\\
&&+P[e_{1}(\widetilde{n_{2}}-1/2)+e_{2}(\widetilde{n_{1}}-1/2)]
\left.+Q(e_{2}+Pe)(\widetilde{n_{2}+Pn}-1/2)
-Qe_{2}(\widetilde{n_{2}}-1/2)\right\}
|e_{3},n_{3}>\nonumber\\
&&=\sum_{e_{3},n_{3}}\delta_{e_{3},e_{1}-Qe}\delta_{n_{3},n_{1}-Qn}\nonumber\\
&&\mbox{exp}-\frac{2\pi
i}{k}P\left[e_{1}(-n-e/2k)-e(n_{1}-Qn+(e_{1}-Qe)/2k-1/2)\right]
|e_{3},n_{3}>\nonumber\\
&&=V
_{*en}^{-Q,P}|e_{1},n_{1}>
\label{A.10}
\end{eqnarray}
It is easy to check that the following linear combination of operators
\begin{equation}
V_{en}^{PQ}=(-)^{a}\sum_{j=0}^{a}(-1)^{j}
\left(\begin{array}{c}
j\\a\end{array}\right)
V_{*e,n-\frac{j}{a}}^{PQ}
\label{A.11}
\end{equation}
satisfies (\ref{A.4}). So we conclude that
\[
V_{en}^{PQ}|e_{1},n_{1}>=
\]
\begin{equation}
(-)^{a}
\sum_{j=0}^{a}(-1)^{j}
\left(\begin{array}{c}j\\a\end{array}\right)
\mbox{exp}\left[2\pi i\frac{Q}{P}(h_{e_{1}+Pe,n_{1}+Pn-\frac{P}{a}j}-
h_{e_{1},n_{1}})\right]|e_{1}+Pe,n_{1}+Pn-\frac{P}{a}j>
\label{A.12}
\end{equation}
If we want to have a standard framing for the torus link, then we
should add an extra factor
\begin{equation}
\mbox{exp}\left(-2\pi i PQh_{en}\right)
\label{A.12***}
\end{equation}
to this
formula. Note  the relation
\begin{equation}
V_{en}^{ap,aq}=\left(V_{en}^{pq}\right)^{a}
\label{A.12*}
\end{equation}
which has an obvious geometric interpretation. Another relation
\begin{equation}
V_{en}^{ap,aq}=(-)^{a-1}\sum_{j=0}^{a}(-1)^{j}
\left(\begin{array}{c}j\\a\end{array}\right)
V_{ae,an-j}^{p,q}
\label{A.12**}
\end{equation}
is due to the cabling properties of the Alexander polynomial.

To calculate the Alexander polynomial of the $(P,Q)$ torus link we
observe that
\begin{eqnarray}
&&V_{en}^{PQ}|0>=-\sum_{l=0}^{\infty}V
_{en}^{PQ}|0,-l>
=(-)^{a+1}\sum_{l=0}^{\infty}\sum_{j=0}^{a}(-1)^{j}
\left(\begin{array}{c}j\\a\end{array}\right)\nonumber\\
&&\mbox{exp}\left[\frac{2\pi i}{k}\frac{Q}{P}Pe(Pn-l-\frac{P}{a}j-1/2)\right]
|Pe,Pn-l-\frac{P}{a}j>
\label{A.14}
\end{eqnarray}
By using  the formula
\begin{equation}
\sum_{k=0}^{j}(-1)^{k}\left(\begin{array}{c}k\\a\end{array}\right)
=(-1)^{j}\left(\begin{array}{c}j\\a-1\end{array}\right)
\label{A.15}
\end{equation}
we can transform eq.~(\ref{A.14}) into the following form
\begin{equation}
V_{en}^{PQ}|0>=(-)^{a+1}
\sum_{j=0}^{a-1}(-1)^{j}\left(\begin{array}{c}j\\a-1\end{array}\right)
\sum_{l=0}^{\frac{P}{a}-1}
\mbox{exp}\left[\frac{2\pi i}{k}Qe(Pn-l-\frac{P}{a}j-1/2)\right]
|Pe,Pn-l-\frac{P}{a}j>
\label{A.16}
\end{equation}
The Alexander polynomial of the unknot carrying a representation
$|Pe,Pn-l-\frac{P}{a}j>$ is equal to $-1/(2i\sin \frac{\pi}{k}Pe)$.
Since
\begin{equation}
\sum_{l=0}^{\frac{P}{a}-1}\mbox{exp}\left(-\frac{2\pi i}{k}Qle\right)=
\frac{1-\mbox{exp}(-\frac{2\pi i}{k}\frac{PQ}{a}e)}{1-\exp(-\frac{2\pi
i}{k}Qe)}
\label{A.17}
\end{equation}
and
\begin{equation}
\sum_{j=0}^{a-1}(-1)^{j}\left(\begin{array}{c}j\\a-1\end{array}\right)
\mbox{exp}\left(-\frac{2\pi i}{k}\frac{PQ}{a}ej\right)=
\left[1-\mbox{exp}\left(-\frac{2\pi i}{k}\frac{Pq}{a}e\right)\right]^{a-1}
\label{A.18}
\end{equation}
then after adding an extra factor~(\ref{A.12***}) to correct the
framing, we get the following expression for the Alexander polynomial
of a torus link:
\begin{equation}
\Delta=(-2i)^{a-2}\frac{(\mbox{sin}\pi PQe/ak)^{a}}
{\mbox{sin}(\pi Pe/k)\mbox{sin}(\pi Qe/k)}
\label{A.19}
\end{equation}
in agreement with  known results.

We derived equation (\ref{A.12}) by using inductive arguments. Now we
present a direct derivation of eqs.~(\ref{A.12}) and
(\ref{A.16}) which also makes contact with some results of
\cite{KS91} and with the Burau matrix
representation of \cite{moody} discussed in \cite{RS91}. We will calculate a
scalar product which is equal to a matrix element of $V
_{en}^{PQ}$ with both indices lowered:
\begin{equation}
<e_{2},n_{2}|V_{en}^{PQ}|e_{1},n_{1}>=
\left[V_{en}^{PQ}\right]_{(e_{2},n_{2})(e_{1},n_{1})}
\label{A.20}
\end{equation}
This scalar product is equal to the invariant of a manifold
$S^{1}\times S^{2}$ obtained by glueing together  two solid tori, one
of which has a Wilson line with representation $(e_{2},n_{2})$ inside,
while the other one has a Wilson line with representation
$(e_{1},n_{1})$ inside and a torus link $(P,Q)$ with representation
$(e,n)$ on its surface.

An $S^{2}$ section of the whole manifold $S^{1}\times S^{2}$ in drawn
in figure 33. As we know, an invariant of the link in such a manifold is
equal to the supertrace of the braiding matrix $B_{Q}$, which shifts the
$(e,n)$ vertices cyclically by $Q$ positions. The supertrace should be taken
over the space of all conformal blocks. A typical conformal block in the
free-field representation is drawn in figure 34. The $n$ screening
operators are integrated along the contours $C_{i_{1}}\ldots
C_{i_{m}}$. Let ${\cal V}_{m}$ denote the
$\left(\begin{array}{c}m\\P\end{array}\right)$-dimensional space of
conformal blocks with $m$ screening operators. The $N$-charge $n_{2}$
in these blocks is fixed by the anomalous charge conservation:
\begin{equation}
n_{2}=m-n_{1}-Pn+1
\label{A.21}
\end{equation}
The action of the braiding matrix $B_{Q}$ consists of shifting the
contours $C_{i}$
\begin{equation}
B_{Q}:\;\;C_{i}\mapsto C_{i+Q}
\label{A.22}
\end{equation}
and multiplicating the conformal blocks by a phase factor
\begin{equation}
\mbox{exp}\left[2\pi i\frac{Q}{P}(h_{e_{2}n_{2}}-h_{e_{1}n_{1}})\right]
\label{A.23}
\end{equation}
Apart from this factor, the matrices $B_{Q}$ ($0\leq Q<P$) form a
group isomorphic to $Z_{P}$.

To simplify our discussion we take the case when $P=p$ and $Q=q$ are
coprime (our arguments can be extended to a general case). The action
of $Z_{p}$ in ${\cal V}_{m}$ ($1\leq m\leq p-1$) splits this space
into the $\frac{(p-1)!}{m!(p-m)!}$ $p$-dimensional invariant
subspaces. The eigenvalues of $B_{q}$ in these subspaces are $p$th
order roots of unity up to the common factor~(\ref{A.23}), therefore
\begin{equation}
\mbox{Str}_{{\cal V}_{m}}B_{q}=0\;\;\;\;\mbox{for}\;\;1\leq m\leq p-1
\end{equation}
However ${\cal V}_{1}$ and ${\cal V}_{p}$ are 1-dimensional
representations of $Z_{p}$, therefore in those spaces the supertrace
of $B_{q}$ should be equal to the factor~(\ref{A.23}). In fact, the
supertrace in ${\cal V}_{p}$ gets an extra factor of $(-1)^{p}$ due to
its fermionic parity, also the operator $B_{q}$ itself gets a factor
$(-1)^{(p-1)q}$, because it permutes fermionic screenings. Since $p$
and $q$ are coprime, the product of both factors is always equal to
$-1$. Thus we see that
\[
<e_{2},n_{2}|V_{en}^{pq}|e_{1},n_{1}>=
\]
\begin{equation}
-\mbox{exp}\left[2\pi i\frac{P}{Q}(h_{n_{2},e_{2}}-h_{e_{1},n_{1}})\right]
(\delta_{-e_{2},e_{1}+pe}\delta_{1-n_{2},n_{1}+pn}-\delta_{-e_{2},e_{1}+pe}
\delta_{1-n_{2},n_{1}+pn-p})
\label{A.24}
\end{equation}
which is equivalent to eq.~(\ref{A.12}) if we put there $a=1$.

Now we turn to the case when $V_{en}^{pq}$ acts on
a vacuum representation $|0>$, i.e. the operator $V_{e_{1}n_{1}}$
is removed from figure 34. Let us first recall that according to
\cite{RS91} the action in ${\cal V}_{1}$ of the
braiding of two neighboring operators $V_{en}$, one of
which is screened (see figure 35), is given by Burau matrices. Thus in
the free-fermion representation of the Alexander polynomial developed
in \cite{KS91}, unscreened operators $V_{en}$
correspond to the strands carrying a vacuum state, while the screened
ones correspond to a one-fermion state flowing along the strand.
Indeed, the screening operators are similar to free fermions: they are
mutually local and fermionic, which means that their permutation
inside a correlator causes a change in the overall sign.

Let us introduce a normalized matrix $B_{q}^{\prime}$:
\begin{equation}
B_{q}^{\prime}=\mbox{exp}\left(-2\pi i\frac{p}{q}h_{pe,pn}\right)B_{q}
\label{A.25}
\end{equation}
whose property is that it acts trivially in ${\cal V}_{0}$. It is easy
to see that the action of $B_{q}^{\prime}$ in ${\cal V}_{m}$ is
isomorphic to its action in $\Lambda^{m}{\cal V}_{1}$ ($\Lambda^{m}$
denotes the $m$th antisymmetric tensor power). Therefore (see
\cite{KS91} for details)
\begin{equation}
\mbox{det}_{{\cal V}_{1}}\left[I-sB_{q}^{\prime}\right]=
\sum_{m=0}^{p}s^{m}\mbox{Str}_{{\cal V}_{m}}B_{q}^{\prime}
\label{A.26}
\end{equation}
here $s$ is a complex variable, $I$ is an identity matrix and
obviously $\mbox{Str}_{{\cal V}_{m}}=(-1)^{m}\mbox{Tr}_{{\cal
V}_{m}}$.

The formula~(\ref{A.26}) requires an important modification. The
screening contours of figure 34 do not produce the valid conformal blocks
when the operator $V_{e_{1},n_{1}}$ is absent, as in this case. The
valid screening contours are those of figure 36: $C_{ij}=C_{i}-C_{j}$.
This means that ${\cal V}_{1}$ should be factored over the
one-dimensional subspace generated by the block with the screening
contour $C=\sum_{i=1}^{p}C_{i}$ and the appropriate factorizations
should also be made in other spaces ${\cal V}_{m}$. The determinant in
eq.~(\ref{A.26}) should therefore be taken over the factor-space
${\cal V}_{1}^{\prime}$, then we will get the supertraces of
$B_{q}^{\prime}$ over the valid conformal block spaces ${\cal
V}_{m}^{\prime}$ in the r.h.s. of eq.~(\ref{A.26}).

A conformal block in ${\cal V}_{1}$ with the screening contour $C$ is
an eigenvector of $B_{q}^{\prime}$ with an eigenvalue equal to $1$, so
a factorization over that vector is equivalent to the regularization
of the braiding determinant for Alexander polynomial considered in
\cite{KS91}. The remaining eigenvalues of
$B_{q}^{\prime}$ in ${\cal V}_{1}$ are the $p$th order roots of unity
$\mbox{exp}(2\pi i l/p)$ ($1\leq l\leq p-1$) up to a phase factor
$\mbox{exp}(\frac{2i\pi}{k}qe)$. So by using the simple algebraic relation
\begin{equation}
\prod_{l=1}^{p-1}[x-\mbox{exp}(2i\pi l/p]=
\frac{\prod_{l=0}^{p-1}(x-\mbox{exp}2i\pi l/p)}{x-1}=
\frac{x^{p}-1}{x-1}=\sum_{l=0}^{p-1}x^{l}
\label{A.27}
\end{equation}
we conclude that
\begin{eqnarray}
\mbox{det}_{{\cal V}_{1}^{\prime}}\left(I-sB_{q}^{\prime}\right)=
\sum_{m=0}^{p-1}\left[s\mbox{exp}(-\frac{2\pi i}{k}qe)\right]^{m}
\label{A.28}\\
\mbox{Str}_{{\cal V}_{m}^{\prime}}=\mbox{exp}\frac{2\pi i}{k}
[qe(pn-1/2)-qme]
\label{A.29}
\end{eqnarray}
Formula~(\ref{A.29}) leads directly to eq.~(\ref{A.16}) taken at
$a=1$.

Finally we briefly comment on a special case when the operator $V
_{en}^{pq}$ acts on a state $|-pe,n_{1}>$. The resulting state
has its $E$-charge equal to zero. In fact, this state is a linear
combination of block representations:
\begin{equation}
V_{en}^{pq}|-pe,n_{1}>=\sum_{n_{2}}A_{n_{2}}|\hat{n}_{2}>
\label{A.30}
\end{equation}
To find the coefficients $A_{n_{2}}$ we calculate the scalar product
$<n_{2}|V_{en}^{pq}|-pe,n_{1}>$, that is, the vertex
$V_{e_{2},n_{2}}$ in figure 34 becomes $V_{n_{2}}$. Note that
\begin{equation}
<n_{2}|V_{en}^{pq}|-pe,n_{1}>=A_{-n_{2}}
\label{A.32}
\end{equation}

The operator $V_{n_{2}}$ has zero $E$-charge, so it is local
with respect to the screening current ${\cal J}$ and it has no poles
in the operator product expansion with that current. Hence the
conformal block corresponding to the contour $C^{\prime}$ in figure 37 is
again equal to zero. It is easy to see ont the other hand that
\begin{equation}
\int_{C^{\prime}}{\cal J}dz=2i\sin
\left(\frac{\pi}{k}e\right)\sum_{i=1}^{p}
\int_{C_{i}}{\cal J}dz
\label{A.33}
\end{equation}
This means that similarly to the previous case, the sum of conformal
blocks corresponding to the contours $C_{i}$ is equal to zero. The
repetition of the arguments that lead us to eq.~(\ref{A.29}) produces
the following result
\begin{equation}
V_{en}^{pq}|-pe,n_{1}>=\mbox{exp}(-2\pi
i\frac{p}{q}h_{-pe,n_{1}})\sum_{m=0}^{p-1}|\widehat{n_{1}+pn-m-1}>
\label{A.34}
\end{equation}
Note that if we assume naively that $|\hat{n}>=|0,n>-|0,n+1>$, then
eq.~(\ref{A.34}) is reduced to eq.~(\ref{A.24}). In fact, the
formula~(\ref{A.34}) requires corrections of order $\epsilon$, but we
will not comment on their derivation here.

\pagebreak

{\bf Figure Captions}

Figure 1: Schematic representation of a "block representation" in $gl(1,1)$
where arrows indicate the action of the fermionic generators. By convention the
state with lowest $n$ number is bosonic $(b)$.

\smallskip

Figure 2: A one dimensional representation $(n)$ must be considered as an
infinite sum of "pesudo typical" representations $(e=0,n')$.

\smallskip

Figure 3: To compute $S$ matrix elements we consider the $U(1,1)$ WZW model on
a torus with a representation $(en)$ running along $\tau$ and insertion of
 $(e_{0}n_{0})$ and its conjugate. A first
 block can be obtained by turning to free field representation
and integrating the screening operator along the indicated dotted contour.

\smallskip

Figure 4, Figure 6: Various possible contours of integration for the screening
 operator.

\smallskip

Figure 5: Set of contours for computing $\zeta_{n'}$

\smallskip

Figure 7: By definition the metric $<\rho_{i}|\rho_{j}>$ is the partition
function of $S^{2}\times S^{1}$ with two parallel Wilson lines carrying
respectively $\rho_{i}$ and $\rho_{j}$.

\smallskip

Figure 8: A Hopf link  made of two linked loops (by convention we take positive
crossings).

\smallskip

Figure 9: Schematic representation of an $S^{2}\times S^{1}$ containing a
representation $(en/I)$. Dotted lines indicate the position of the regulators.

\smallskip

Figure 10: Schematic representation of an $S^{2}\times S^{1}$ containing a
representation $(en/II)$.

\smallskip

Figure 11: The scalar product of two representations of  type $I$. A non
vanishing result is obtained as $\epsilon\rightarrow 0$.

\smallskip

Figure 12: The scalar product of a representation  of type $I$ and a
representation of type $II$. Again a non vanishing result is obtained as
$\epsilon\rightarrow 0$.

\smallskip

Figure 13: The scalar product of two representations of type $II$. A detached
loop is obtained, that leads to a vanishing invariant.

\smallskip

Figure 14: The regularized value of $S_{nn'}$ corresponds to a Hopf link with
two loops carrying $(n)$ and $(n')$ and connected by a pair of dotted lines. It
amounts to a single dotted loop, with invariant going as $1/\epsilon$.

\smallskip

Figure 15: A situation with  two unlinked circles is equivalent to a situation
where a loop carrying a one dimensional representation connects them.

\smallskip

Figure 16: Schematic representation of fusion $I.I$, $I.II$, $II.II$.

\smallskip

Figure 17: The factorization formula for two disconnected links can be
established if one connects them with a pair of dotted lines.

\smallskip

Figure 18: Verlinde formula expresses the consistency in calculating the
invariant of this link.

\smallskip

Figure 19: When $\rho_{i}$ is a four dimensional indecomposable block,
the Hilbert space obtained by cutting the corresponding loop by an $S^{2}$ has
dimension greater than one, and the factorization formula cannot be used.

\smallskip

Figure 20: A double of the trefoil.

\smallskip

Figure 21: Double where stands carry the same representation and have
opposite orientations.

\smallskip

Figure 22: A twisted double of the trefoil.

\smallskip

Figure 23: A three cable of the trefoil.

\smallskip

Figure 24: Surgery on the indicated system of loops produces the manifold
$X_{h}\times S^{1}$.

\smallskip

Figure 25: Surgery on this link produces $S^{1}\times S^{1}\times S^{1}$.

\smallskip

Figure 26: After surgery on loop 1, $S^{2}\times S^{1}$ is obtained.

\smallskip

Figure 27: The partition function of figure 26 is evaluated by factorization
and leads to considering figure 27.

\smallskip

Figure 28: The case where $\rho=(\widehat{n})$ is studied by considering
instead a pair of Wilson lines with opposite orientations that carry $(en/I)$.

\smallskip

Figure 29: Cap 1 extracts the conformal block $|1>$ without logarithms.

\smallskip

Figure 30: Cap 2 extracts a  conformal block $|2>$ with logarithms.

\smallskip

Figure 31, 32: Link configurations used in the computation of $Tr{\cal M}$.

\smallskip

Figure 33: A section $S^{2}$ of the manifold $S^{2}\times S^{1}$ obtained by
glueing two solid tori, one of which has a Wilson line with representation
$(e_{2}n_{2})$ inside, while the other has a Wilson line with representation
$(e_{1}n_{1})$ inside and o torus link $(p,q)$ with representation $(en)$ on
its surace.

\smallskip

Figure 34: A typical conformal block in the free field representation, where
doted lines represent screening contours.

\smallskip

Figure 35: Braiding of two vertex operators, one of which is screened, is
represented by the Burau matrix \cite{RS91}.

\smallskip

Figure 36: When the operator $V_{e_{1}n_{1}}$ is absent, the correct screening
contours are the $C_{ij}=C_{i}-C_{j}$.

\smallskip

Figure 37: The conformal block corresponding to the contour $C'$ vanishes.

\pagebreak

\end{document}